%% file: 00_manuscript.tex
\documentclass[a4paper,fleqn]{cas-dc}

\usepackage[authoryear]{natbib}

\def\tsc#1{\csdef{#1}{\textsc{\lowercase{#1}}\xspace}}
\tsc{WGM}
\tsc{QE}
\tsc{EP}
\tsc{PMS}
\tsc{BEC}
\tsc{DE}

\begin{document}
\let\WriteBookmarks\relax
\def\floatpagepagefraction{1}
\def\textpagefraction{.001}
\shorttitle{}
\shortauthors{L. Malenica et~al.}

\title [mode = title]{Towards improved understanding of spontaneous imbibition into dry porous media using pore-scale direct numerical simulations}                     



\author[1]{Luka Malenica}[orcid=0000-0002-8309-588X]
\cormark[1]
\ead{lmalenica@ethz.ch}

\credit{Conceptualization, Investigation, Methodology, Software, Validation, Visualization, Writing – original draft, Writing – review \& editing}

\affiliation[1]{organization={Institute for Building Materials (IfB), ETH Zurich},
                city={Zurich},
                postcode={8093}, 
                state={},
                country={Switzerland}}

\author[1]{Zhidong Zhang}[]

\credit{Conceptualization, Formal analysis, Investigation, Validation, Writing – review \& editing}

\author[1]{Ueli Angst}[]

\credit{Conceptualization, Funding acquisition, Investigation, Resources, Supervision, Writing – review \& editing}

\cortext[cor1]{Corresponding author}


\begin{abstract}
Traditional approaches to mathematically describe spontaneous imbibition are usually based on either macro-scale models, such as Richards equation, or simplified pore-scale models, such as the bundle of capillary tubes (BCTM) or pore-network modeling (PNM). It is well known that such models cannot provide full microscopic details of the multiphase flow processes and that many pore-scale mechanisms still lack proper mathematical descriptions. To improve the predictive capabilities of traditional models, a fundamental understanding of pore-scale dynamics is needed. 
The focus of this paper is obtaining detailed insight and consistent explanation of particular processes during capillary-controlled water imbibition into dry porous media. We use two-dimensional model geometries and perform fully dynamic volume-of-fluid based direct numerical simulations of air-water multiphase flow at the pore-scale, to study processes that generally are not considered in traditional models. More specifically, we investigate differences between converging and diverging geometries, dynamic pressure and meniscus reconfiguration during pore-filling events, and the influence of inertia and pore size on imbibition dynamics and the occurrence of capillary barriers. Furthermore, we perform a detailed comparison between non-interacting and interacting BCTM and study the impact of the narrow contractions on imbibition dynamics and the trapping of the non-wetting phase. Obtained knowledge can be used to improve predictive models, which are broadly relevant considering the importance of spontaneous imbibition in many different natural and industrial processes.
\end{abstract}





\maketitle

\sloppy
\input{01_Introduction}

\input{02_Methodology}

\input{03_Results}

\input{04_Implications}

\input{05_Conclusions}

\printcredits

\section*{Acknowledgement}

This work was supported by the Swiss National Science Foundation (SNSF, grant no. 196919) and the European Research Council (ERC) under the European Union's Horizon 2020 research and innovation program (grant agreement no. 848794).

\bibliographystyle{cas-model2-names}

\bibliography{06_Bibliography}


\end{document}

%% file: 01_Introduction.tex
\section{Introduction}\label{sec_introduction}

Spontaneous imbibition is a process in which a wetting fluid is spontaneously drawn into porous media displacing a nonwetting fluid due to capillary action \citep{morrow2001recovery}. Such spontaneous invasion of the wetting phase is important in many scientific and engineering applications, such as hydrocarbon recovery \citep{andersen2014model}, $CO_2$ sequestration \citep{scanziani2020dynamics}, water percolation through the unsaturated zone \citep{malenicaGroundwaterFlowModeling2018}, moisture transport in building materials \citep{zhang2020dual}, water uptake by plants \citep{wang2021reduced}, design of waterproof clothing \citep{ezzatneshan2020study}, inkjet printing \citep{aslannejad2021liquid} or micro heat pipes \citep{mooney2022capillary}, just to name a few. 

Spontaneous imbibition is a complex process governed by many different properties, particularly pore space structure and the pore size distribution, wetting properties (fluid–fluid and porous media–fluid surface interactions), initial water saturation, and the density/viscosity ratios of involved fluids. 
Despite a significant amount of research done over the last few decades \citep{adler1988multiphase,gray1991unsaturated,mason2013developments,li2019spontaneous},
significant gaps between simulations, laboratory results and field-scale measurements prevail in complete understanding of flow phenomena in porous media. Thus, a deeper fundamental understanding of spontaneous imbibition processes is essential and has a large potential to improve our
predictive capabilities.

Traditional approaches to mathematically describe spontaneous imbibition are usually based on simplified models. At the macro-scale, the Darcy law for saturated flow \citep{gotovac2021control} is extended by introducing the relative permeability concept \citep{bear1987effective} and used to study the multiphase flow in porous media \citep{whitaker1986flow_II}. Historically, two independent research aspects led to the development of continuum-scale multiphase flow models. In hydrology, soil scientists were mostly interested in water movement through unsaturated soil. For such applications, neglecting the air phase was often reasonable and led to the development of the well-known Richards equation \citep{richrds1931capillary} and many applications afterward \citep{malenica2019numerical,zha2019review,zhang2020modeling}. On the other hand, in petroleum engineering, both phases (usually liquid-liquid system) need to be considered, which led to the development of two-phase flow equations for porous media \citep{muskat1936flow}. In both cases, similar assumptions were used. First, the volume averaging technique was applied to derive continuum equations, meaning that flow variables, such as capillary pressure, velocity or saturation, represent averaged values over some representative elementary volume (REV). Moreover, multiphase flow is a fundamentally dynamic process where one phase replaces the other, thus the inherent assumption of static configuration of fluid arrangements inside REV is not actually valid \citep{blunt_multiphase_2017}. Furthermore, the dependence on the flow rate and viscous coupling between phases were usually ignored which are important physical factors affecting multiphase flow in porous media. 
Accuracy of the macro-scale approaches highly depends on the relative permeability curves which generally assume that macroscopic capillary pressure is a function of saturation only. However, it is well known that the interaction of capillary and viscous forces at the pore-scale is highly complex and that different pore-scale properties, such as contact angle, pore morphology, viscosity ratio and interfacial area, among others, play important roles \citep{armstrong2012linking}. Different attempts to understand and to include various individual effects into relative permeability curves were published \citep{hassanizadeh1993thermodynamic,reeves1996functional,porter2009lattice,liu2011hysteresis,fan2019comprehensive,guo2022role}; however, many oversimplifications of the pore-scale physics still persist and the validity of relative permeability curves is still an open question.

Parallel to the development of macro-scale models, pore-scale modeling has evolved \citep{golparvar2018comprehensive}. Such models aim to provide more detailed microscopic insight into multiphase flow without empirical macroscopic properties such as relative permeability \citep{blunt_multiphase_2017}. Pore-scale studies started by work of Lucas \citep{lucasUeberZeitgesetzKapillaren1918a} and Washburn \citep{washburnDynamicsCapillaryFlow1921} where co-current imbibition of water displacing air in a single capillary tube was studied and led to the development of the well-known Lucas-Washburn equation (LWE). Since then, many theoretical and experimental studies led to modifications and extensions of LWE to account for different effects, such as gravity, inertia, tortuosity, dynamic contact angle, different cross-section shapes and axial variations, among others \citep{mason2013developments,caiLucasWashburnEquationBasedModeling2021}.
LWE still represents a fundamental equation for modeling imbibition in a single capillary tube, however, it is well known that there are many challenges when applying it to model realistic porous media. Hence, the bundle of capillary tubes models (BCTM) were developed \citep{yuster1951theoretical,bartleyRelativePermeabilityAnalysis1999,dullien2012porous}, extending the basic LWE approach to better represent realistic complex pore spaces. The main idea was to represent a porous medium as a system of parallel well-defined tubes with varying sizes. Initially, independent capillaries without fluid transfer between tubes were employed. In such models, capillaries with larger diameters are filled first, which however contradicts the behavior observed in real porous media. Thus, interacting capillary bundle tube models with cross-flow between capillaries were developed \citep{dong2005immiscible,dong2006immiscible,wang2008fluid}. 
Interacting BCTM enabled additional insight into the fundamental mechanism of spontaneous imbibition and are still extensively used and under development \citep{li2017crossflow, ashraf2020spontaneous,ashraf2023spontaneous}. {However, in addition to simplified equations, their main drawback is over-simplified geometry representation which, with advances in high-resolution imaging and image processing techniques \citep{kaestner2008imaging,blunt2013pore,prodanovic2023digital,mohamed2023scale,zhang2022dynamic}, it is becoming even more important for modern modeling tools to account for realistic geometrical features of the porous media.}

To capture the complex topology of real pore structures more appropriately, the pore-network models (PNM) were developed \citep{fatt1956network,bakke19973,blunt2001flow}. In PNM, multiphase flow is modeled in a pore-space represented as a network of connected pore bodies and pore throats. Initially, quasistatic PNM similar to invasion-percolation theory \citep{berkowitz1993percolation} were developed. Such models were used to simulate capillary pressure equilibrium states, which has shown appropriate for predicting capillary-dominated systems \citep{celia1995recent,blunt2001flow}. To account for viscous forces and transient processes of multiphase flow between equilibrium states, dynamic PNM were developed \citep{al2005dynamic,joekar2012analysis,qin2019dynamic}. PNM have undergone significant developments and nowadays can accommodate irregular lattices, wetting layer flow, wettability heterogeneity, as well as a variety of different physical processes \citep{xiong2016review,foroughi2020pore,zhao2022dynamic,foroughi2023incorporation}. PNM are extensively used due to their computational efficiency and ability to construct pore-space networks directly from micro-CT images. However, network elements (pores and throats) of real pore spaces are generally approximated with idealized objects, such as spheres and cylinders. In addition to simplifications of the pore-space geometry, different flow assumptions relying on invasion-percolation theory sacrifice the accuracy of multiphase flow description as a trade-off for computational efficiency \citep{pavuluri_towards_2020}. Nonetheless, various relevant pore-scale processes, such as interface reconfigurations, pore-filling rules or trapping mechanisms are still not fully understood and lack adequate mathematical description within the PNM framework.

On the other hand, direct numerical simulations (DNS), often called direct pore-scale modeling (DPSM) within porous media applications, is an approach that does not require macroscopic transport properties or assumptions about pore-scale mechanisms like above described models. All variables are directly computed at the pore-scale level enabling fundamental insight into the microscopic behavior of pore-scale dynamics by solving full governing equations. DNS can handle geometries of arbitrary complexities, thus it can be applied directly on digital twins of real porous media (e.g. acquired through micro-CT or FIB-SEM tomography) without significant simplifications of pore-space geometries. While there are many applications of DNS for studying single phase flow \citep{molins2012investigation,starchenko2016three,ramanuj2020characteristics,ramanuj2022macrovoid}, it was only in the last ten years that DNS started to be applied for studying multiphase flow in porous media \citep{raeini2012modelling,liu2016multiphase,lunati2016special,maes_direct_2018,soulaine2021computational}. Different numerical approaches have been adjusted for pore-scale simulations of multiphase flow \citep{zhao2019comprehensive}, such as lattice Boltzmann method (LBM) \citep{liu_influence_2021}, smoothed particle hydrodynamics (SPH) \citep{tartakovsky2016smoothed}, level set method (LSM) \citep{friis2019pore}, volume of fluid (VOF) method \citep{shams2018numerical} and phase field method \citep{frank2018direct}. While DNS simulations of multiphase flow are computationally demanding and require sophisticated numerical algorithms, their main strength is that they can provide insights into variables and processes that conventional models are not able to or that are too difficult to measure experimentally. Thus, they can help us obtain deeper fundamental understanding and ensure that the pore-scale physics is accurately scaled to maintain accuracy and relevance or macroscale models which, together with development of numerical methods \citep{malenica2021full,liem2022adaptive,wu2023new}, should lead to the development of improved macro-scale and predictive pore-scale models \citep{ferrari2013direct,raeini2018generalized,menke_upscaling_2021,giudici2023representation}. 

This work aims to improve the understanding of spontaneous imbibition mechanisms using direct numerical simulations. Most DNS applications so far were related to hydrocarbon recovery, i.e. water-oil systems. As one of our particular interests is understanding the capillary absorption of water in building materials, such as concrete, stone or brick, we consider the immiscible flow of air and water in this work. However, the significance of the air-water system extends beyond building materials and applies to various fields, such as rainfall infiltration or water uptake by plants, for example. We believe understanding simple systems where particular processes can be investigated in detail is essential before tackling more complex problems and geometries. Thus, in this work, we focus on relatively simple two-dimensional geometries and investigate different effects, such as pore- and throat-filling events and the importance of inertia (at different scales) on the occurrence of capillary barriers during spontaneous imbibition of water in dry porous media. Furthermore, we perform detailed DNS comparisons between classical (non-interacting) and interacting BCTM and demonstrate some new insights. Finally, additional reasons and profound insights into why water imbibes faster through smaller capillaries than in large ones (even without cross-flow between capillaries like in interacting BCTM) are demonstrated together with its effect on the air trapping mechanism. We believe that such a basic understanding of flow physics is of essential importance for developing reliable large-scale models.

%% file: 02_Methodology.tex
\section{Methodology}\label{sec_methodology}
\subsection{Governing equations for two-phase flow at the pore-scale}
\label{subsec_fullEQ}

\subsubsection{Equations of motion}
\label{subsec_NS}

In this work, we consider the incompressible flow of two immiscible Newtonian phases. Such flow can be described by mass conservation and one-fluid formulation of Navier-Stokes equations (NSE) for momentum conservation given by \citep{tryggvason_direct_2011}:

\begin{equation}
    \nabla \cdot \mathbf{u}=0
    \label{eq:MassConserv}
\end{equation}

\begin{equation}
    \frac{\partial \rho 
 \mathbf{u}}{\partial t} + \nabla \cdot (\rho \mathbf{u} \mathbf{u}) = - \nabla p + \nabla \cdot \{\mu [\nabla \mathbf{u} + (\nabla \mathbf{u})^T]\} +\rho \mathbf{g} + \mathbf{f}_{st}
 \label{eq:MomentumConserv}
\end{equation}

where $\mathbf{u}$ is velocity vector, $t$ is time, $p$ is pressure, $\rho$ is density, $\mu$ is viscosity, $\mathbf{g}$ is gravity vector and $\mathbf{f}_{st}$ is surface tension force. For constant interfacial tension, the surface tension force can be described by the Continuous Surface Force (CSF) model \citep{brackbill_continuum_1992}:

\begin{equation}
    \mathbf{f}_{st}=\sigma \kappa \mathbf{n} \delta_{\Gamma}
    \label{eq:CSF}
\end{equation}

where $\sigma$ is interfacial tension, $\kappa$ is interface curvature, $\mathbf{n}$ is normal vector to fluid/fluid interface and $\delta$ is Dirac delta function located at the interface $\Gamma$.

Equations (\ref{eq:MassConserv})-(\ref{eq:MomentumConserv}) are valid for the whole flow field even in the case when fluid properties, such as density and viscosity, change discontinuously. However, different fluids must be identified. In this work, the algebraic Volume of Fluid (VOF) interface capturing approach \citep{hirt_volume_1981} is used to distinguish between different fluids using a fixed computational grid.

\subsubsection{Volume of fluid (VOF) method}
\label{subsec_VOF}

 In the VOF approach, the interface is tracked using an indicator function $\alpha$ which represents the volume fraction of one of the fluids in each grid cell. The indicator function takes value $\alpha=1$ for the first fluid and $\alpha=0$ for the second fluid, while the interface is located at the position where fluid properties change abruptly. 
Using the indicator function field, the density and viscosity are defined using their single-filed values:

\begin{equation}
    \rho=\alpha \rho_1 + (1-\alpha) \rho_2
    \label{eq:vof_rho}
\end{equation}

\begin{equation}
    \mu=\alpha \mu_1 + (1-\alpha) \mu_2
    \label{eq:vof_mu}
\end{equation}

where $\rho_i$ and $\mu_i$ are density and viscosity of phase $i$. Similarly, the velocity and pressure fields can be defined as:

\begin{equation}
    \mathbf{u}=\alpha \mathbf{u}_1 + (1-\alpha) \mathbf{u}_2
    \label{eq:vof_u}
\end{equation}

\begin{equation}
    p=\alpha p_1 + (1-\alpha) p_2
    \label{eq:vof_p}
\end{equation}

The additional transport equation needs to be solved to advect indicator function (i.e. fluid/fluid interface):

\begin{equation}
    \frac{\partial \alpha}{\partial t} + \nabla \cdot (\alpha \mathbf{u}) = 0
 \label{eq:VOF}
\end{equation}

 In the VOF method, the normal vector needed to calculate surface tension force (\ref{eq:CSF}) is defined by:

\begin{equation}
    \mathbf{n}=\frac{\nabla \alpha}{||\nabla \alpha||}
 \label{eq:vof_normal}
\end{equation}

while the curvature can be calculated by taking the divergence of the normal field as: 

\begin{equation}
    \kappa=\nabla \cdot \mathbf{n} = \nabla \cdot \biggl( \frac{\nabla \alpha}{||\nabla \alpha||} \biggr)
 \label{eq:vof_curvature}
\end{equation}

Wetting properties of the porous media are defined by specifying a normal vector to enforce equilibrium contact angle $\theta$ at the fluid/fluid/solid contact line:

\begin{equation}
    \mathbf{n}=\mathbf{n}_s cos \theta + \mathbf{t}_s sin \theta
    \label{eq:normal_BC}
\end{equation}

where $n_s$ and $t_s$ are the normal and tangent unit vectors to the solid surface, respectively.

In this work, \textit{interGCFoam} solver implemented in GeoChemFoam code is used for the numerical solution of governing equations presented in this section. GeoChemFoam is a state-of-the-art OpenFOAM-based numerical solver particularly developed for direct numerical simulations at pore-scale and has been applied to a variety of problems, such as upscaling single-phase flow in a micro-porous carbonate \citep{menke_upscaling_2021}, wettability alteration during low-salinity flooding \citep{maes_direct_2018}, dissolution of $CO_2$ bubbles \citep{maesDirectPoreScaleModelling2018, maes_unified_2020}, mineral dissolution \citep{maes_improved_2022,menke2023channeling}, species transfer across fluid interfaces \citep{maes_new_2018}, modeling flow and heat transfer \citep{maes_geochemfoam_2022} and multiphase reactive transport \citep{maes_geochemfoam_2021} using micro-CT images of real porous media.

\textit{interGCFoam} uses a collocated finite volume grid with a predictor-corrector strategy based on the Pressure Implicit Splitting Operator (PISO) algorithm \citep{issa_solution_1986} and explicit coupling between NSE and interface advection equation (\ref{eq:VOF}). It implements different numerical procedures to improve the accuracy and stability of the classical VOF approach which is needed for demanding pore-scale multiphase flow modeling. For more details regarding numerical procedures implemented in GeoChemFoam, we refer to \citep{maes2021geochemfoam, maes_geochemfoam_2021}.

\subsection{Materials, numerical setups and fluid properties}
\label{subsec_system}

Our research interest is understanding water absorption into cement-based materials, such as mortar or concrete \citep{zhangDifferentAnomaliesTwostage2024}, with a special interest in air-water distribution at the steel-concrete interface as this is highly relevant for steel corrosion in reinforced concrete \citep{angst_steelconcrete_2017,malenicaDirectNumericalModelling2024}. The microstructure of cementitious materials is highly heterogeneous, with pore sizes ranging from nanometer to millimeter scale. Even though very fine (nm scale) porosity exists in such materials, larger capillary pores (in the µm range) are assumed to dominate water absorption properties \citep{dongCharacterizationComparisonCapillary2017}. Thus, most of the presented numerical setups are based on capillaries with radii between 1 and 20 µm. An additional example, ranging radius from 0.1 to 100 µm, was used to investigate the capillary size effects outside the mentioned range. Even though our primary interest is in cementitious porous materials, numerical setups are created to represent specific geometrical features of general porous media and are not in any way restricted to a single particular porous material.

Note that the term radius is used throughout the manuscript for simplicity. It actually represents the half-width of the capillary since the flow through a 2D capillary is equivalent to flow between two infinitely long parallel plates where curvature in the perpendicular direction is zero.

Moreover, in all examples, we consider air and water systems with the following fluid properties: $\rho_{w}=1000$ [kg/m$^3$], $\rho_{a}=1.225$ [kg/m$^3$], $\mu_{w}=10^{-3}$ [Pa s], $\mu_{a}=1.81 \cdot 10^{-5}$ [Pa s] and $\sigma=0.072$ [N/m], where subscripts $w$ and $a$ denote water and air phase, respectively. Gravitational acceleration $g$ [m/s$^2$] is set to zero because its effects on the flow dynamics compared to the capillary forces are negligible due to the small capillaries and domain lengths considered in this work. The contact angle value is assumed to be constant and set to $\theta$ = 5°, as our focus is on the air-water system under strongly water-wet conditions.

%% file: 03_Results.tex
\section{Results and discussion}

\label{sec_Results}

In this section, we present the numerical results together with discussions highlighting processes that are still poorly understood and generally not accounted for in simplified pore-scale models, such as PNM and BCTM.
The verification examples, where DNS results are compared to analytical solutions (under both static and dynamic conditions), can be found in Supplementary Information.

\subsection{Understanding effects of abrupt changes in geometry on multiphase flow}

\subsubsection{Pore-filling vs throat-filling}
\label{subsec_PoreVSThroatFilling}

Firstly, we examine imbibition dynamics considering sharp changes in capillary shape, such as diverging and converging geometries. The former is a representative case of the pore-filling process, while the latter is of the throat-filling process. 
In general, most simplified pore-scale models, such as PNM or BCTM, do not account for what happens in the transition zone (the zone where geometry changes abruptly) but rely on the capillary pressure values and geometrical features (i.e., the radius of the capillary) before and after the transition zone. Here, we use simple geometries to demonstrate that dynamic effects occurring in the transition zone are important and can play a significant role in overall imbibition dynamics. 

\begin{figure*}
\centering{\includegraphics[width=0.9\textwidth]{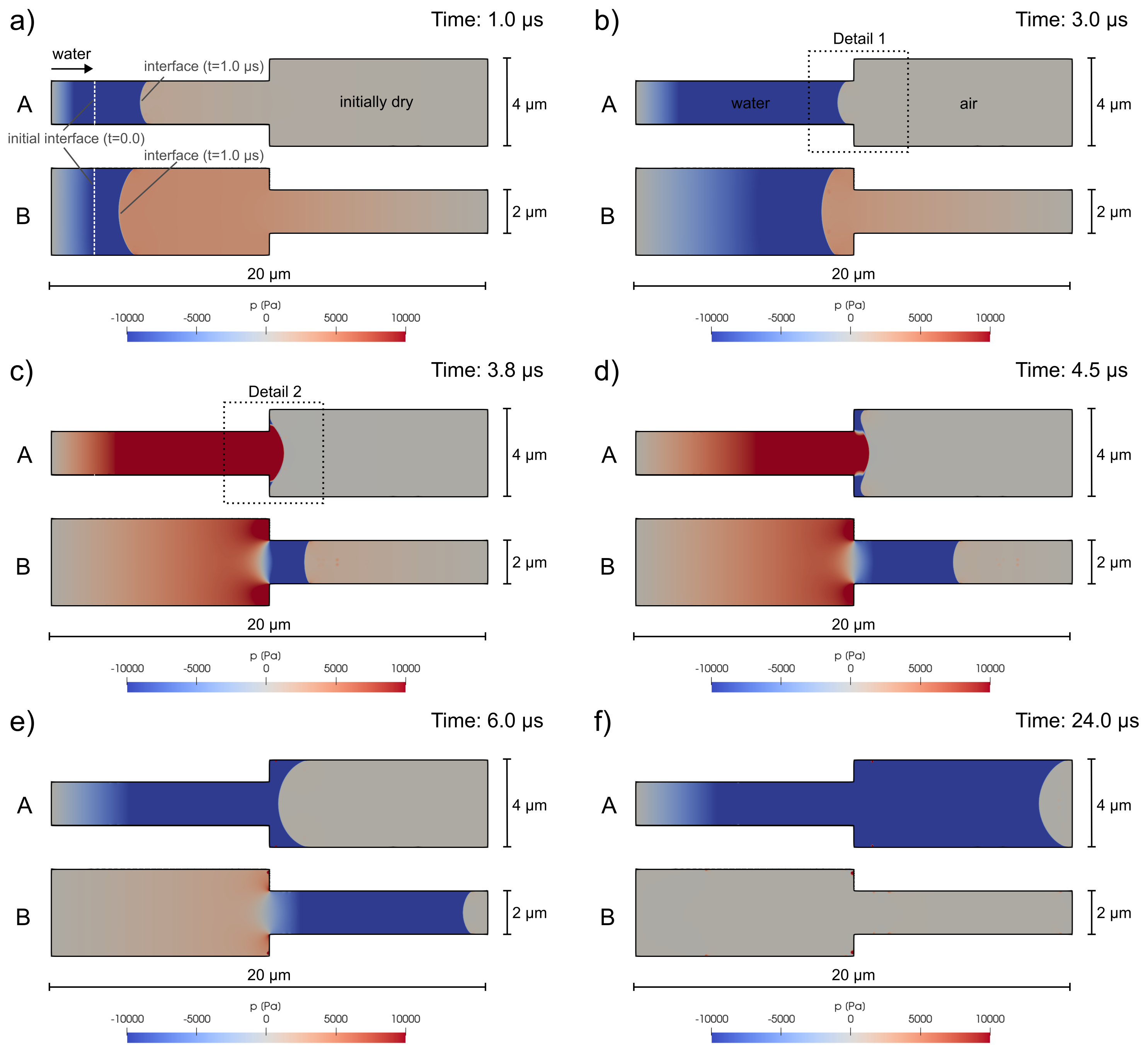}}
  \caption{Pressure distribution for two cases of abrupt pore geometry changes distinguishing between negative (blue) and positive (red) pressure zones and illustrating that pore filling (A) is a significantly slower process than throat filling (B). Dashed lines at time t=1.0 µs represent the air-water interface at t=0 (initial condition). The complete animation showing the full process can be found in the Supplementary Materials.}\label{fig_sGeom_06_A_vs_B}
\end{figure*}

Figure \ref{fig_sGeom_06_A_vs_B} shows the pressure distribution at six different times, where case A represents pore-filling while case B represents a throat-filling event. The radius of the narrow (throat) and wide (pore body) part of the capillary are chosen 1 µm and 2 µm, respectively. The domain's left and right ends are free inlet and outlet boundaries, respectively, where the fluid (either liquid or gas) can freely enter or leave the domain. The domain's top and bottom boundaries represent walls with a specified contact angle value ($\theta$ = 5°). Dashed white lines on Figure \ref{fig_sGeom_06_A_vs_B}a represent (initially flat) air-water interface at t=0 (i.e., initial condition). As we consider spontaneous imbibition, the pressure boundary conditions at the inlet and outlet are set to zero ($p=0$), so initially, there is no driving force in the time $t=0$. However, because of interfacial tension between air and water and the wettability of water with solid boundaries (i.e., $cos(\theta)\neq0$) at the walls, this will lead to the formation of curved meniscus producing the non-zero capillary pressure so the capillary driven flow will occur.

\begin{figure}
\centering{\includegraphics[width=1.0\columnwidth]
{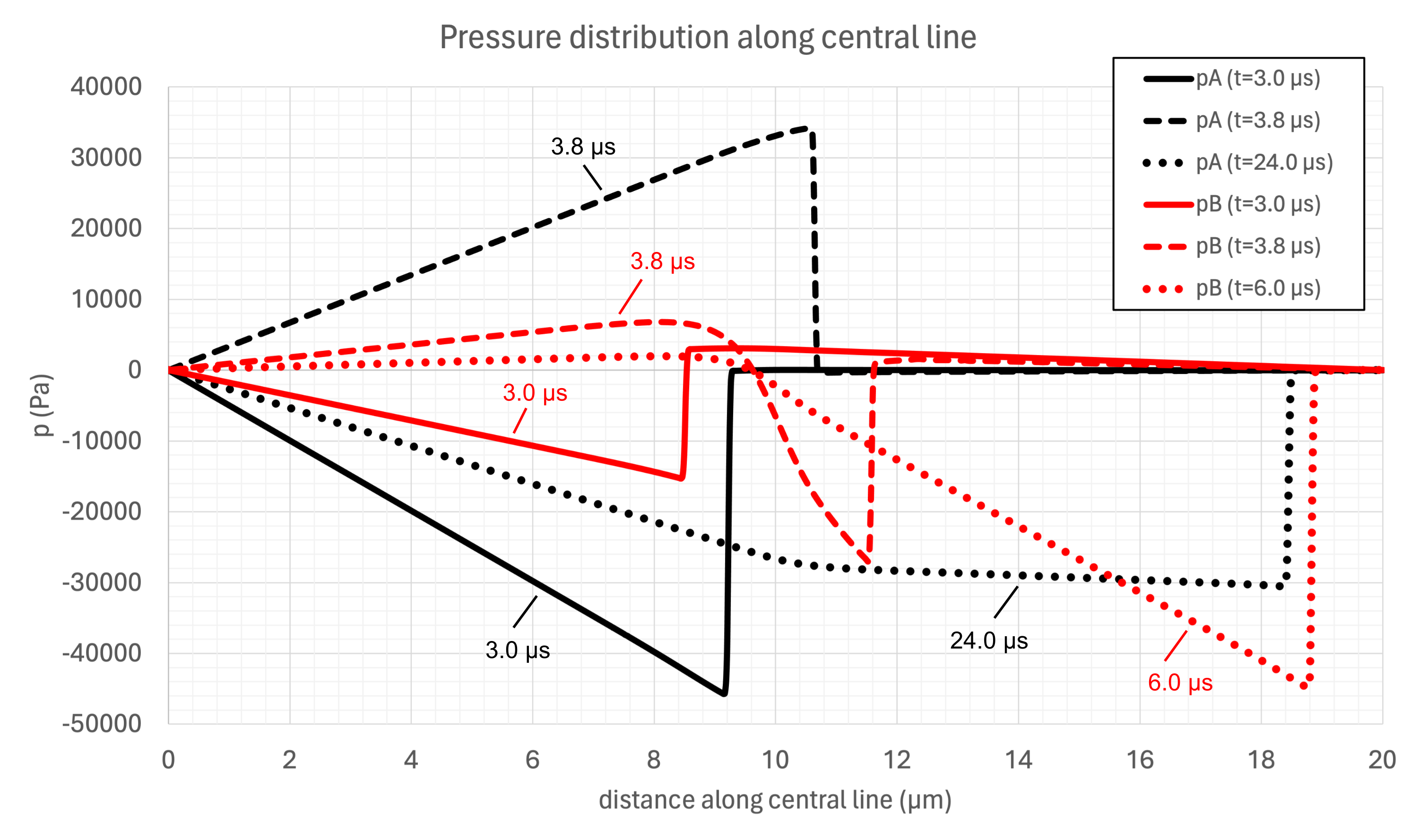}}
  \caption{Pressure distribution along the central lines for chosen times from Figure \ref{fig_sGeom_06_A_vs_B}. pA represents pore-filling while pB represents throat-filling. The complete animation showing the full process can be found in the Supplementary Materials.}\label{fig_sGeom_06_A_vs_B_pLine}
\end{figure}

A color map for pressure distribution is chosen in an equal range for both negative and positive values to clearly distinguish between under-pressure (blue) and over-pressure (red) fluid zones. In addition to 2D pressure plots,  corresponding 1D pressure lines, representing pressure distribution along the central lines, are plotted (for 3 chosen times for each case) in Figure \ref{fig_sGeom_06_A_vs_B_pLine} where pressure jumps represent both (dynamic) capillary pressure $P_c$ values as well as the position of the interface.

In the beginning (before reaching the transition zone), the water imbibition is faster in the narrow capillary (case A) as visible in Figures \ref{fig_sGeom_06_A_vs_B}a, \ref{fig_sGeom_06_A_vs_B}b and Figure \ref{fig_sGeom_06_A_vs_B_pLine} (black solid vs red solid line). This is due to inertia effects as the larger capillary (case B) contains a greater mass of water, which needs more time to be accelerated at the beginning of the simulation. If the distance to the transition zone was significantly longer, the meniscus in case B would (at some point) overtake the meniscus distance of case A, and imbibition in the wider capillary would be faster, as described by the basic Lucas-Washburn equation (LWE).

By looking at the pressure fields before reaching the transition zone (Figures \ref{fig_sGeom_06_A_vs_B}a, \ref{fig_sGeom_06_A_vs_B}b and \ref{fig_sGeom_06_A_vs_B_pLine} (t=3.0µs)), it is visible that the water phase is experiencing negative pressure values (under-pressure) for both cases, which is generally expected for the wetting phase during the imbibition process. On the other hand, the air phase needs positive pressure downstream of the interface, as a pressure gradient is needed for air (being "pushed" by advancing meniscus) to flow toward the outlet boundary (where pressure is set to zero as the boundary condition for spontaneous imbibition). For case A, over-pressure values in the air phase are low and hardly distinguished from zero pressure values, while for case B, they are larger and non-negligible due to two main reasons: 1) larger volumetric discharge and 2) larger viscous dissipation due to a narrower radius downstream of the transition zone. However, even for case B, over-pressure values in the air phase are significantly lower than (absolute) under-pressure values in the water phase because air has significantly lower viscosity than water (water is approximately 55 times more viscous than air).
 
Particular focus lies on understanding processes once the interface reaches the transition zone. The throat-filling (case B) is an energetically favorable and rapid process, and fast penetration of water from the pore body to the throat can be observed (Figure \ref{fig_sGeom_06_A_vs_B}b-e). Multiple factors complementing each other are responsible for this. First, the meniscus transition from a wide pore body to a narrow throat is a quick process where the interface reaches throat walls almost instantaneously, allowing for fast reconfiguration to a stable interface with a large positive (concave) curvature in a narrower throat. 
Moreover, the reduction in the cross-sectional area will lead to a temporary build-up of (over-) pressure in the water phase, leading to the (temporary) meniscus speed-up even if the total flow rate decreases (due to the reduction of geometry). This over-pressure is expected to be released over time.
Further, the larger curvature in the throat leads to larger capillary pressure (driving force), while the viscous dissipation in the upstream part is now reduced as the friction in the pore body is practically negligible due to the larger cross-sectional area compared to the narrow throat. This is visible by looking at evolution of the pressure curve slopes (red lines in Figure \ref{fig_sGeom_06_A_vs_B_pLine}), where pressure gradients are directly related to viscous dissipation. Eventually (Figure \ref{fig_sGeom_06_A_vs_B}f), once meniscus leaves the domain, there won't be driving force anymore, and after friction dissipates residual inertial effects, the flow rate will go to zero and uniform pressure distribution (p=0) is expected. 

On the other hand, pore filling (case A) is an energetically highly unfavorable process as the air-water interface needs to bulge from a narrow throat into a wide pore body, resulting in negative capillary pressure, which is acting as an opposing force to the water imbibition as visible in Figure \ref{fig_sGeom_06_A_vs_B}c. Note that negative capillary pressure, in our case, implies positive water pressure and vice versa due to the definition of capillary pressure ($P_c=P_a-P_w$). The air-water interface must reach the pore body walls first (Figure \ref{fig_sGeom_06_A_vs_B}d) before it can obtain positive curvature again (\ref{fig_sGeom_06_A_vs_B}e). After it reaches pore body walls and establishes positive capillary pressure (leading to under-pressure of water phase once again), the imbibition process is significantly slower as the driving force (capillary pressure $P_c$) is smaller than it was inside the throat (larger radius means smaller $P_c$), while the main viscous dissipation is still occurring in the throat which is now filled by water. The pressure line at time t=24µs (Figure \ref{fig_sGeom_06_A_vs_B_pLine}) is now divided into two slopes upstream of the interface. The first part related to the flow in the throat has a larger slope (large energy dissipation due to small radius), while the pressure loss in the pore body part is almost negligible as most of the energy is dissipated into the narrow (throat) part of the system.

As demonstrated in this example, throat filling is generally a rapid process and is not expected to slow down the imbibition process. On the other hand, pore filling is a significantly slower process. In this case, filling the diverging geometry is more than four times slower than filling the same geometry from the opposite direction. Moreover, the actual pore-filling process with such sharp changes in geometry (when fluid needs to bulge from the throat into the pore body with negative capillary pressure) can have a significant effect on overall imbibition and dynamic effects happening in the transition zone between throat and pore body are generally not accounted in classical imbibition modeling approaches. The following section looks into the pore-filling process in more detail.

\subsubsection{Detailed insight into dynamic pressure and meniscus reconfiguration during the pore-filling process}
\label{subsec_Detail}

\begin{figure*}
\centerline{\includegraphics[width=0.85\textwidth]{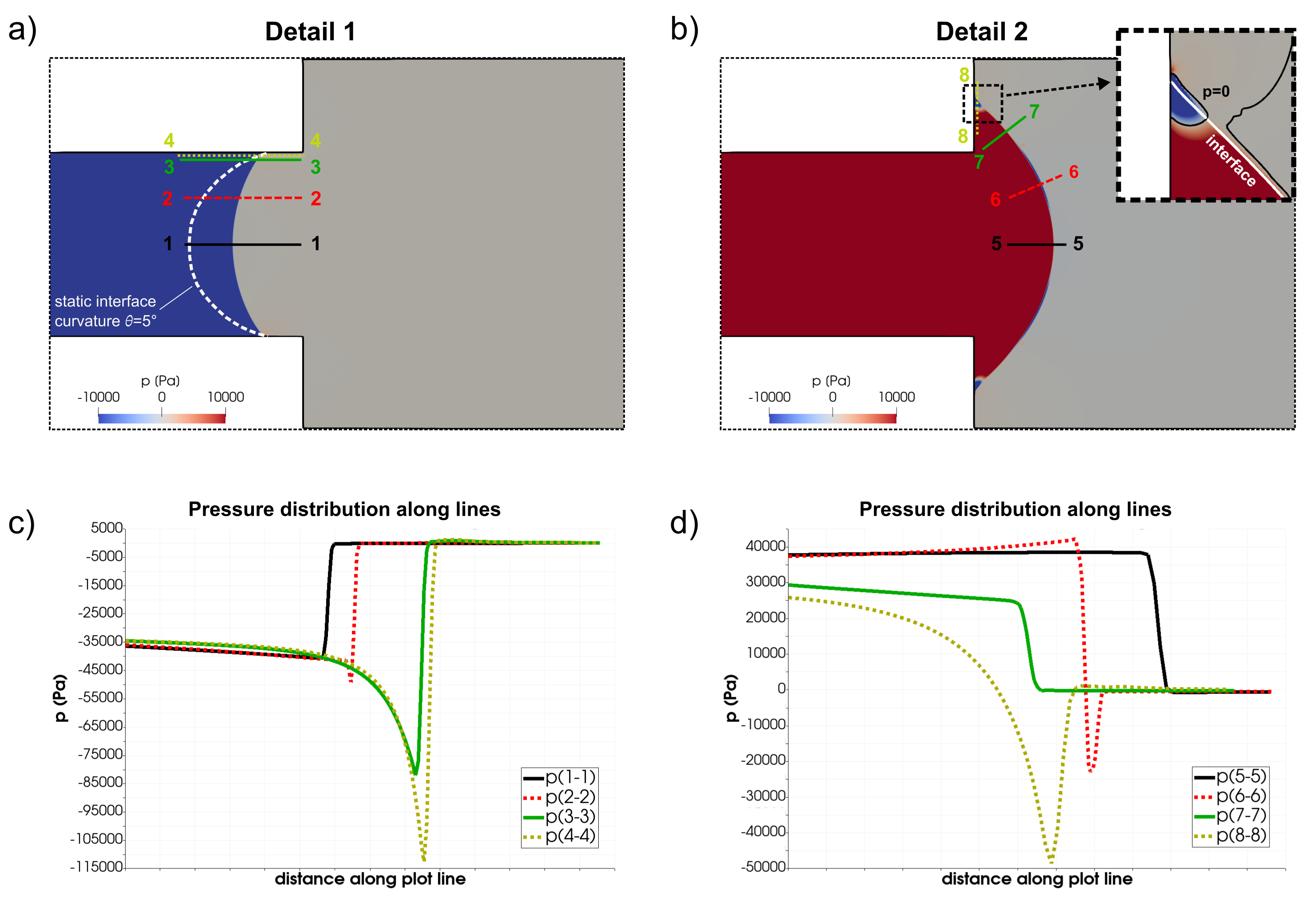}}
  \caption{Insight into pressure field near the air-water interface during the pore-filling event showing that dynamic capillary pressure varies significantly across the interface. Figures a) and b) are zoom-in details denoted in Figures \ref{fig_sGeom_06_A_vs_B}b and \ref{fig_sGeom_06_A_vs_B}c while figures c) and d) represent line pressure plots across colored lines denoted in a) and b), respectively. White dashed line on \ref{fig_sGeom_06_detail_p_pL}a represents (constant curvature) shape of the meniscus under static conditions for contact angle $\theta=5$°.}\label{fig_sGeom_06_detail_p_pL}
\end{figure*}

A more detailed insight into (dynamic) pressure distribution during pore-filling events is performed by zooming in the two details marked on Figures \ref{fig_sGeom_06_A_vs_B}b and \ref{fig_sGeom_06_A_vs_B}c and shown on Figure \ref{fig_sGeom_06_detail_p_pL}. Again, the same (equal) pressure range is used as in Figure \ref{fig_sGeom_06_A_vs_B}, while insight into specific capillary pressure values and pressure ranges can be obtained from pressure line plots below each zoom-in detail.

Figure \ref{fig_sGeom_06_detail_p_pL}a shows a meniscus in a narrow throat just before reaching the transition zone. A few observations can be stated here. First, strong water-wetting conditions were assumed by setting contact angle $\theta=5$°. This means that the meniscus shape should be close to the shape of the semicircle (which would be expected in the case of perfectly wetting fluid), as it is marked with a white dashed line. However, such curvature is only expected under static conditions, and it is clear that the (dynamic) meniscus obtained from numerical simulations is significantly less curved. It can also be stated that curvature over the meniscus is not constant, which is in accordance with different capillary pressure values at different positions, as seen in line pressure plots in Figure \ref{fig_sGeom_06_detail_p_pL}c. Most pore-scale models, such as PNM and BCTM, generally rely on capillary pressure values calculated by the Young-Laplace (Y-L) equation. In this case, for given parameters (surface tension, contact angle, and throat radius), the capillary pressure value calculated by Y-L equation would be 71726 Pa. However, it is clear that (dynamic) capillary pressure values are not constant across the interface. The lowest value is in the middle of the meniscus (approximately half of the value predicted by the Y-L equation), and it increases by approaching the capillary walls as visible by looking at pressure plots across lines 2-2, 3-3 and 4-4. The presented numerical results indicate that local capillary pressure near the wall can be significantly higher than one obtained by the Y-L equation (line 4-4).

In the VOF method, the contact angle is incorporated as boundary conditions, where the normal vector at the fluid/fluid/solid contact line is adjusted to enforce the specified contact angle value. This means that the algorithm should enforce the expected curvature locally (near the solid walls). In the case of static configuration, this will lead to constant capillary pressure and constant curvature (defined by contact angle value) across the whole meniscus (to the accuracy of the numerical algorithm). However, during dynamic processes, this is not necessarily true and leads to a lower apparent contact angle and unequal capillary pressure distribution over the meniscus. The opposite situation (with higher apparent contact angle) is expected during a drainage process, which is in accordance with contact angle and capillary pressure hysteresis, where the relation $\theta_R \leq \theta \leq \theta_A$ holds. Here $\theta_R$, $\theta$, and $\theta_A$ stand for receding (drainage), static, and advancing (imbibition) apparent contact angles, respectively. Thus, even before reaching the transition zone, dynamic capillary pressure is not necessarily constant across the meniscus and has different values from the static (Young-Laplace) pressure. Even though the constant (equilibrium) contact angle value was used for simulations, the dynamic effects that produce apparent contact angle repercussions observed in reality are, to some extent, accounted for within the VOF framework.

\begin{figure*}
\centerline{\includegraphics[width=0.9\textwidth]{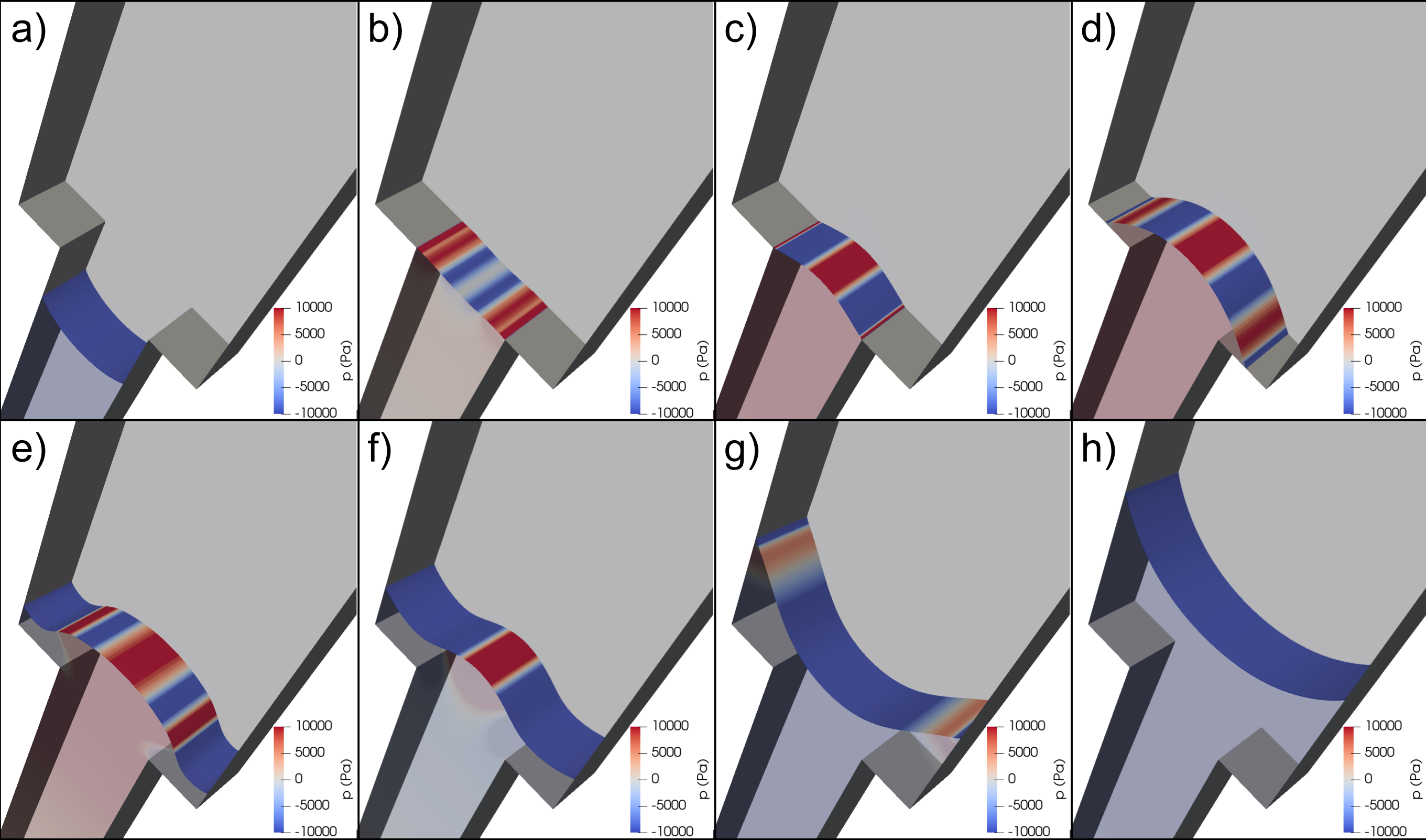}}
  \caption{Time evolution (a-h) of the complex meniscus reconfigurations during the pore-filling event. Colors on the interface contour distinguish between different under- and over-pressure zones at the interface. The complete animation showing the full process can be found in the Supplementary Materials. }\label{fig_sGeom_06_interfaceView}
\end{figure*}

Figure \ref{fig_sGeom_06_detail_p_pL}b shows Detail 2, where the meniscus is bulging inside the pore body with opposite (convex) curvature and where water pressure has undergone changes from under-pressure to over-pressure (i.e., negative (apparent) capillary pressure has been established). This negative capillary pressure acts as an opposing force to the flow and slows down the imbibition process. However, the VOF results again show large capillary pressure variations across the interface, where some parts of the interface are now (locally) experiencing even under-pressure values. Lines 5-5 and 7-7 in Figure \ref{fig_sGeom_06_detail_p_pL}d have strictly positive pressure values. However, the pressure values over lines 6-6 and 8-8 clearly show the existence of negative pressure values near the interface. The negative (under-) pressure (i.e., positive capillary pressure) near the solid boundary (line 8-8) is explained by the local suction effects by locally establishing concave curvature (due to water-wetting properties of the solid boundaries). The zoom-in view in Figure \ref{fig_sGeom_06_detail_p_pL}b shows magnified local pressure distribution together with interface position (white line) and zero pressure contours (black lines). It is visible that this pressure drop occurs only locally near the interface. Even though it seems that the under-pressure zone extends slightly to the air phase, actually, the whole pressure drop occurs only throughout the water phase. This is due to the (diffusive) nature of the VOF method, where the interface cannot fully accurately be represented with a single contour line, as shown here. However, this representation is a good indicator of the interface zone since, on this scale, a fully sharp assumption of an interface is reasonable (in reality, the thickness of the air-water interface is estimated to be below 1 nanometer).
This assumption is also supported by the shape of the pressure line 8-8, where (relatively sharp) pressure jump (positive capillary pressure) is clearly visible and believed to occur exactly over the air-water interface. 
Moreover, pressure line plot 6-6 shows an even more complex pressure distribution. Approaching the interface, the pressure seems to experience a slight increase before the rapid drop to negative pressure values, which is again followed by a jump to (approximately) zero pressure values (as expected in the air phase). At this moment, we cannot give a comprehensive explanation regarding such large local variations and the existence of under-pressure values at this specific location of the interface. For a better insight into pressure variations across the interface, we refer to the animation related to Figure \ref{fig_sGeom_06_interfaceView}, which is discussed in the following.

Figures \ref{fig_sGeom_06_detail_p_pL}b and \ref{fig_sGeom_06_detail_p_pL}d indicate that the single meniscus experiences both positive and negative pressure zones simultaneously. Additional insight into complex meniscus reconfigurations during the pore-filling process is shown in Figure \ref{fig_sGeom_06_interfaceView}, where pressure colors are used to distinguish between such zones directly at the interface. 
Initially, Figure \ref{fig_sGeom_06_interfaceView}a shows a meniscus in a narrow throat before reaching the transition zone, and as explained previously, the water phase and interface are experiencing under-pressure values. 
Further, in Figure \ref{fig_sGeom_06_interfaceView}b, the air-water interface has reached the transition zone, and complex interface reconfiguration with different pressure zones starts occurring. After this, the interface begins bulging inside the pore body, leading to meniscus curvature changes as shown in Figure \ref{fig_sGeom_06_interfaceView}c. This change from concave to convex curvature changes water pressure from under-pressure to over-pressure; however, as discussed previously, some zones near the interfaces have negative pressure values (blue zones). At this point, the interface is temporarily pinned, and once the pressure (due to carried momentum) increases enough, the de-pinning will occur, and the meniscus will start advancing on the vertical walls (Figure \ref{fig_sGeom_06_interfaceView}d). Even the wettability effects produced localized pressure drop and positive capillary pressure next to the wall (as described above), a significant portion of the surface area experiences negative capillary pressure, and the central part of the bulge is expected to start retreating. This will spread water toward vertical pore-body walls and temporally accelerate the imbibition process. 

Once the water reaches corners and touches pore body walls (Figure \ref{fig_sGeom_06_interfaceView}e), meniscus reconfiguration toward positive meniscus curvature will start. During this process, a complex meniscus shape experiences both positive (near horizontal walls where water as wetting fluid is driving imbibition forward) and negative capillary pressure areas (mostly in the central part of the bulge, which is responsible for water flow towards corners but also for temporally reversing flow direction in the throat) can be seen as shown in Figure \ref{fig_sGeom_06_interfaceView}f. This will be followed with the meniscus transition toward positive capillary pressure (Figure \ref{fig_sGeom_06_interfaceView}g) until the situation without over-pressure in the water phase happens once again (Figure \ref{fig_sGeom_06_interfaceView}h), as it was the case before meniscus reached transition zone (Figure \ref{fig_sGeom_06_interfaceView}a). Note that interface oscillations in the pore-body (after passing the transition zone) can take some time until the meniscus reaches steady-state (dynamic) curvature again. 

Here, it was shown that interface reconfigurations during the pore-filling process involve complex multiphase flow dynamics with a range of capillary pressure values, including both positive and negative values. In the following sections, the influence of inertia and absolute pore system size will be investigated.

\subsubsection{Importance of inertia during pore filling}
\label{subsec_Inertia}

While considering both single-phase and multiphase flow in low-permeability porous media, the inertia effects are usually neglected because of low (averaged) velocities in porous media. This is the case for most pore-scale models, such as BCTM or PNM, as well as Darcy law-based models (e.g., Richards equation), which predict the diffusion-type behavior of the flow. In this work we emphasize that inertia, even on small scales (such as micron and sub-micron size), can significantly affect filling time and be a determining factor if pore filling will occur. 

Recently, pore-scale understanding of pore-filling events started gathering attention and the importance of pore-space geometry and inertia started to be recognized. \citet{ferrari_inertial_2014} have studied inertial effects on meniscus reconfiguration and surface-to-kinetic energy transformation of pore-filling events during forced imbibition. They recognized that the meniscus reconfigurations have their own time scales and local velocities which can be several orders of magnitude larger than injection velocities. It was shown that inertia effects in complex media influence the pore-filling order, leading to different invasion patterns and fluid-fluid distributions. \citet{zacharoudiou_pore-filling_2017} performed both experimental and numerical investigations of the pore-filling sequence by considering spontaneous imbibition and recognized that the geometry of pore space is the main deciding factor in the displacement pathways. Both works called into question the simple pore-filling rules based on capillary force. However, in all of their test cases, the wetting phase has always invaded the pore body.   

Recently \citet{pavuluri_towards_2020} investigated the concept of capillary barriers \citep{wu2016two,hu_investigation_2022} at pore-scale and found that geometric characteristics of the pore space, contact angle and the fluid properties play an important role during pore filling. They observed the situation where larger pore bodies are getting invaded first, and the smaller pore bodies, in some cases, never get invaded due to the capillary barrier effect, which is the opposite behavior predicted by conventional pore network models. 
They analytically defined \textit{critical} contact angle ($\theta_c$), which depends on the slope of the wall boundary in the transition zone between the throat and the pore body. If the contact angle of the wetting phase is lower than the critical contact angle ($\theta < \theta_c$), the wetting phase will imbibe the pore body during spontaneous imbibition, while for ($\theta \geq \theta_c$) capillary barrier should occur. However, they also recognized that inertia can assist the wetting phase to overcome the capillary barrier zones.  
They considered the influences of different contact angles on fluid-fluid systems with the same density and viscosity and used steady-state meniscus velocity as the initial condition. Note that in our case steady-state velocity cannot exist (even for a single capillary with constant radius) due to differences in viscosities between air and water. 

More recently, \citet{liu2021pore} simulated capillary barrier effects in pore-throat structures using LBM. They developed an efficient depth-averaged quasi-3D color-gradient LBM model by assuming a constant contact angle in the depth direction. They confirmed this assumption with full 3D LBM simulations for two different contact angle values; however, the influence of inertia on overcoming capillary barriers was not investigated. 

\begin{figure*}
\centering{\includegraphics[width=0.9\textwidth]{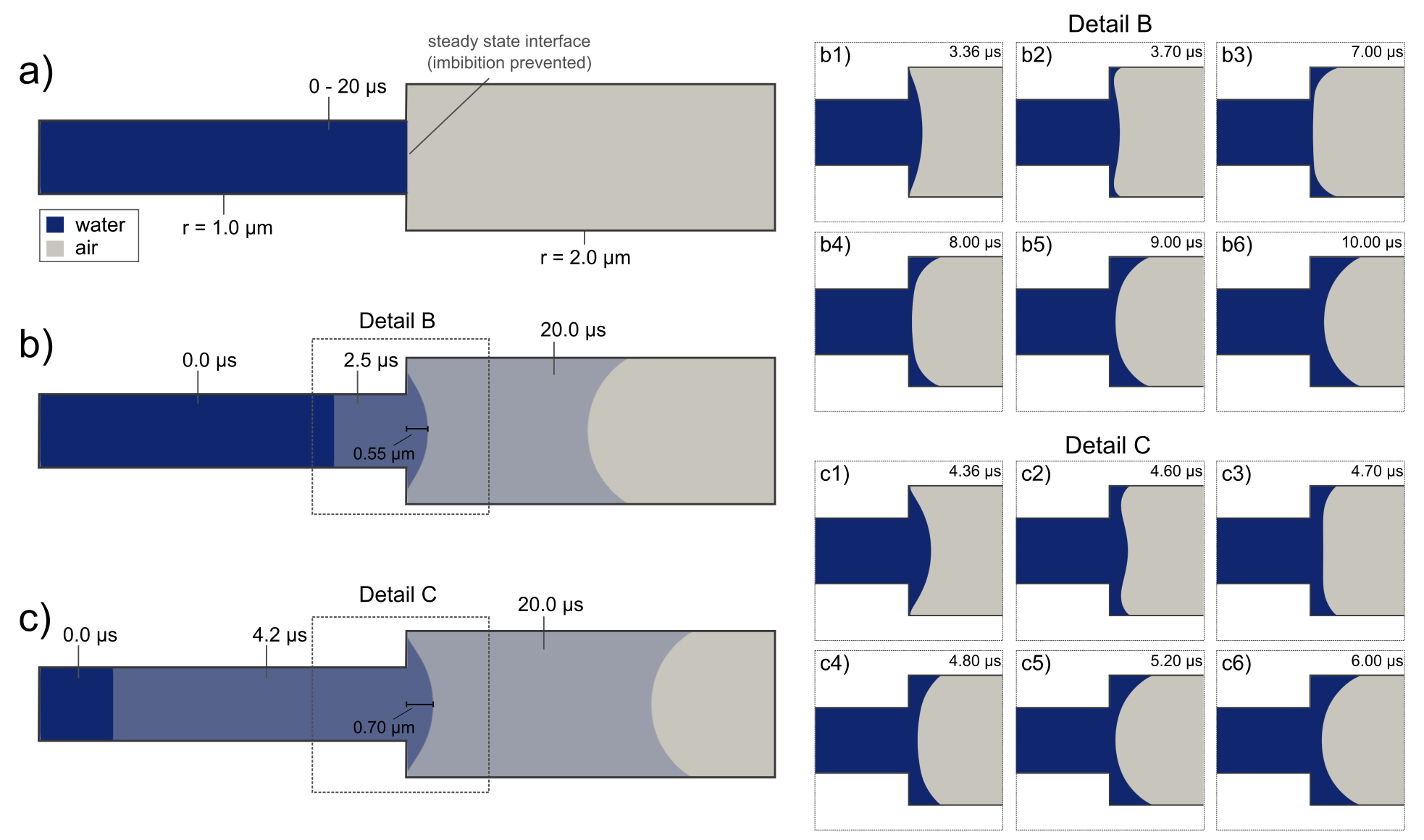}}
\caption{Influence of inertia on pore filling process. The figures a), b) and c) show the air-water distribution at 3 different times for each of 3 different cases, where different shades of blue color show time evolution. For case a), spontaneous imbibition is prevented by the occurrence of a capillary barrier due to positioning the initial interface directly in the transition zone. For cases b) and c), the momentum carried by water is large enough to overcome the capillary barrier, and water will invade the pore body. The larger momentum of case c) will eventually lead to faster filling of the pore body than case b) even if the initial interface was initially further away. Maximum momentum is characterized by the values of the maximum bulge (0.55 and 0.70 µm). Comparing all 3 cases, the
pore filling occurs in the opposite sequence (or not at all) compared to the initial imbibition distance because of inertia effects in the transition zone. Details B and C show intermediate times from when the meniscus reaches horizontal walls until stable curvature is achieved. Dimensions of the systems are the same as in Figure \ref{fig_sGeom_06_A_vs_B} (i.e., throat radius is set to 1.0 µm). The complete animation showing the full process can be found in the Supplementary Materials.}\label{fig_sGeom_06_Inertia}
\end{figure*}

In this work, we investigate how inertia influences pore-filling dynamics and whether a capillary barrier will occur during water imbibition into dry porous media with sharp changes in geometry. 
The same pore-filling geometry as in the previous example is used, i.e., the pore body-to-throat radius ratio is 2, and the throat has a radius of 1µm. The importance of inertia is investigated by considering three different initial conditions as shown in Figure \ref{fig_sGeom_06_Inertia}. 

We first consider the case \textit{a} where the initial condition is positioned directly in the throat-pore transition zone. In the previous section, we have seen that pore filling with sharp geometry changes, such as the presence of corners, requires bulging of the fluid from the throat to the pore body with negative capillary pressure. Here, the initial condition is a flat meniscus which corresponds to zero capillary pressure. Due to the wetting effects, we know that a curved meniscus is preferentially formed and capillary-driven flow will occur in case of constant geometry. However, as we have a sharp increase in the capillary radius here, it is energetically not favorable for the meniscus to advance as it should form negative capillary pressure, i.e. opposite curvature. During spontaneous imbibition, there is no external force that would create over-pressure upstream of the interface and "push" the meniscus into the pore body. Thus, for the case \textit{a}, a flat meniscus corresponding to zero capillary pressure is a stable configuration and pore filling will not occur. This phenomenon is known as the capillary barrier \citep{pavuluri_towards_2020}. Here, it is worth mentioning that pore filling can eventually occur during numerical simulations if the simulation is carried out for a long time (particularly if coarse mesh is used). This behavior is not physical. Rather, it is related to numerical diffusion and numerical instabilities known as spurious (or parasitic) currents \citep{pavuluri_spontaneous_2018} which are advecting small oscillations in indicator function (\ref{eq:VOF}) and eventually can accumulate over longer simulation periods to produce the non-physical pore filling. Thus, it is important to be aware of such behavior during simulations. Computational remedies are finer spatial and temporal discretization (mesh size and time step) and a more appropriate interface-capturing approach (e.g., geometric VOF, interface-tracking approaches, etc.).

For the remaining two cases (\textit{b} and \textit{c}), pore-filling will happen as momentum carried by water (when it reaches the transition zone) is large enough to overcome the capillary barrier. For the case \textit{b}, the interface is initially closer to the transition zone and, once formed, the meniscus will reach the transition zone before the case \textit{c} due to the shorter distance. However, the meniscus velocity when it reaches the transition zone will be lower than that of the case \textit{c} due to the shorter acceleration length (as here, initial acceleration still plays a significant role). This will lead to the pore body of case \textit{c} eventually filling faster than the of the case \textit{b}. The reason for this is explained as follows. As mentioned, water must bulge into the pore body with negative capillary pressure during the pore-filling event for the considered geometry. The difference in momentum between the two cases will lead to different maximum bulges into the pore body (0.55 µm vs 0.70 µm) between \textit{b} and \textit{c} cases, respectively. Once the maximum bulge is achieved, the meniscus will tend to retreat because the negative capillary pressure acts as an opposing force and wants to "push" the meniscus back, which will also spread the water toward pore-body walls. A larger maximum bulge means a larger volume of water inside the pore body, so it will reach the horizontal walls (corners) of the pore body faster. Once it reaches corners, meniscus reconfiguration will tend to happen so positive capillary pressure can be established again. However, if the volume of water is insufficient, the non-uniform curvature of the meniscus will be established, as seen in detail B of Figure \ref{fig_sGeom_06_Inertia} (b3). This leads to a less curved interface in the central part of the meniscus meaning that the total driving force (integrated value of capillary pressure $P_c$) is lower than it will be after the meniscus advances further into the pore body (compare meniscus curvature between b3 and b6 case on Figure \ref{fig_sGeom_06_Inertia}). For the case \textit{c}, the process of touching the pore-body walls and forming uniform curvature will be relatively faster due to the larger volume of water produced by the larger bulge into the pore body, which is directly related to the momentum of water carried once the meniscus reaches the throat-pore transition zone. This can be seen by comparing the time needed to establish "stable" curvature from the moment water touches pore-body walls (i.e., times between b1 (c1) and b5 (c5)), where this process is almost 7 times slower for case \textit{b}. Additionally, it is worth mentioning that the time since maximum bulge is achieved till water touches the walls of the pore body is also significantly slower (0.86 µs vs 0.16 µs) for case \textit{b}.

To conclude, for the presented geometry, pore filling occurs in the opposite sequence (or not at all) compared to the initial imbibition distance. The main reason for such behavior is inertia, which is often neglected when modeling multiphase flow in porous media.

\subsubsection{Size effect during pore filling}
\label{subsec_Size}

\begin{figure}
\centering{\includegraphics[width=1.0\columnwidth]
{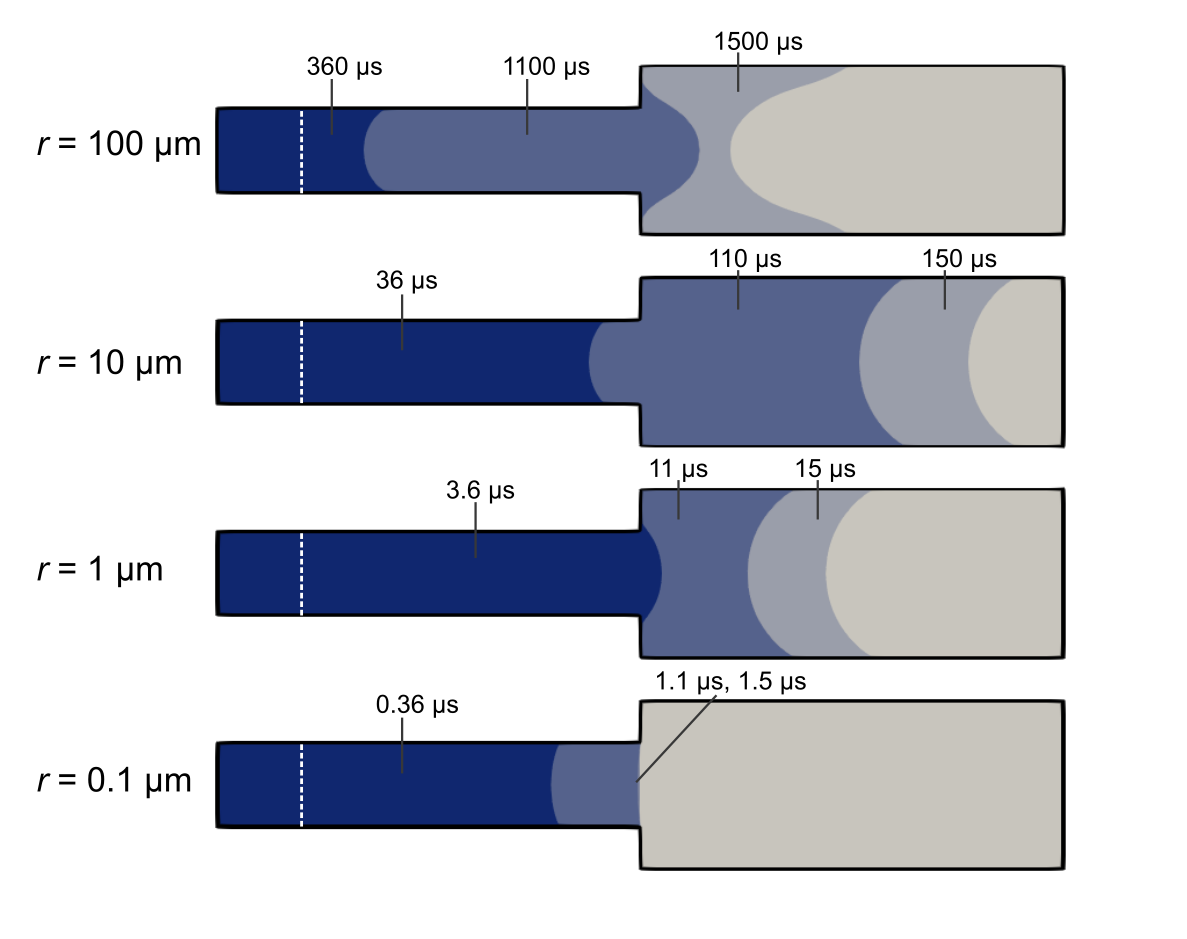}}
\caption{Influence of the spatial scales on the pore-filling process. The Figure shows air-water distributions for 3 different times for each of 4 different throat radii. For the case of r=0.1 µm, results at the 1.1 µs and 1.5 µs overlap, as the interface does not advance inside the pore body. This case demonstrates that capillary barriers are more likely to occur on smaller spatial scales, which contradicts general pore-filling rules in PNM that smaller pores are invaded first due to larger capillary pressure. Dashed lines represent the air-water interface at t=0 (initial condition), while dimensions are scaled by factor 1/r for visualization purposes. The complete animation showing the full process can be found in the Supplementary Materials.}\label{fig_sGeom_06_SizeEffect}
\end{figure}

As a final example using a single capillary with sharp changes in the cross-section, we consider size effects, i.e., the same geometry shape on different spatial scales. The throat radius is varied from 0.1 µm to 100 µm with a factor of 10, while the pore-throat ratio is fixed and kept at 2. Figure \ref{fig_sGeom_06_SizeEffect} shows the air-water distributions for 4 different sizes, where results are scaled in space with a factor of $1/r$, and $r$ represents the capillary radius (half-width in 2D) of the throat.

Dashed lines in Figure \ref{fig_sGeom_06_SizeEffect} represent the initial condition (flat interface). If the throat distances were sufficiently long, we would expect that the meniscus velocity would increase with the increasing radius of the throat, just as predicted by the Lucas-Washburn equation. Thus, in larger throats, the meniscus would reach the transition zone faster (relatively comparing, i.e., scaled in both space and time by factor 1/r). However, for a chosen system, the initial acceleration of the liquid plays a role, so at the larger scales, more time is needed for water to accelerate from initially stagnant liquid (due to the larger associated mass). In this case, we see that the meniscus in a capillary with $r=$ 1 µm reaches the pore body first (relatively comparing all cases), meaning that inertia effects related to initial water acceleration are becoming negligible on smaller scales than this. 
Also note that on a larger scale, meniscus deformation will be more pronounced, and it will take longer until the steady meniscus shape with steady curvature is established, whereas on small scales, this occurs rapidly. 

Once the meniscus reaches the transition zone, a larger throat radius means larger momentum, so the capillary barrier will be easier to overcome (as demonstrated in the previous example). Here the most important observation is that on the smallest scale ($r=$ 0.1 µm and lower) pore-filling will not happen for a given configuration. Such behavior is often overlooked as smaller pores are assumed to be invaded first (and faster) due to larger capillary pressure. However, here we show that inertia effects play a crucial role and are often the deciding factor if pore-body filling of very small pore bodies will happen.

This has a significant effect on spontaneous imbibition in many porous materials where many small dry pores will not be invaded at all if the geometrical configurations (e.g., small pores surrounded by the network of the even smaller capillaries) are such that it is not energetically favorable for the meniscus to invade the pore body. The presented results agree with \citet{pavuluri_towards_2020}, where simple pore-filling rules used in PNM were called into question.

\subsection{System of capillary tubes}
\label{subsec_SystemOfCapillaryTubes}

\subsubsection{Interacting capillaries}
\label{subsec_InteractinCapillaries}

As discussed in the Introduction, in the classical bundle of capillary tubes (BCTM) approach, the porous medium is represented as a set of separated tubes, each of them modeled by the Lucas-Washburn equation. Such an approach predicts the fastest imbibition in the capillary with the largest radius which contradicts observed behavior in real porous media \citep{bico2003precursors,ashraf2017spontaneous,ashraf2019capillary}. To account for complex fluid interaction occurring in porous media, interacting BCTM were developed \citep{dong2006immiscible}.

In general, interacting BCTM assume pressure equilibrium between capillaries containing the same fluid. This further presumes that the entire fluid transfer takes place only at the meniscus position and neglects any pressure drop between tubes. However, this assumption of pressure equilibrium cannot be physically correct as the pressure difference is needed for the flow between capillaries to occur. Recently, Ashraf et al. \citep{ashraf2018spontaneous} performed VOF simulations where they showed that pressure difference between tubes does exist, but both pressure drop and fluid transfer occur only in the vicinity of the fluid/fluid interface. 
However, in their work, they assumed that two capillaries are continuously connected which could be interpreted as a single capillary setup with a complex (but constant) cross-section.  Such a setup is not a fully valid representation of general porous media where, for example, solid grains act as no-flow boundaries interrupting cross-flow continuity. Thus, in this work, we investigate a setup where capillaries are connected only at discrete locations.  Moreover, \citet{ashraf2018spontaneous} considered that both phases have the same viscosity and density which led to the same pressure gradient before the first and after the second meniscus. In this work, we consider the multiphase flow of water and air, where significant differences in density and viscosity play a role.



\begin{table}[!htbp]
\caption{Dimensions (diameters and lengths in µm) specified in Figures \ref{fig_sGeom_10_0050} and \ref{fig_sGeom_10_throats_01}.}
\centering
\begin{tabular}{ |c|c|c|c|c|c|c|  }
\hline
$D_1$ & $D_2$ & $D_3$ & $L_1$ & $L_2$ & $L_3$ & $L_4$\\
 \hline
12 & 40 & 8 & 100 & 200 & 300 & 400 \\
 \hline \hline
$L_5$ & $L_6$ & $L_7$ & $L_8$ & $L_9$ & $L_{10}$ & $L_{11}$\\
 \hline
52 & 96 & 100 & 60 & 400 & 40 & 120 \\
 \hline
\end{tabular}
\label{tab:sGeom_10_dimensions}
\end{table}

\begin{figure*}
\centering{\includegraphics[width=0.95\textwidth]{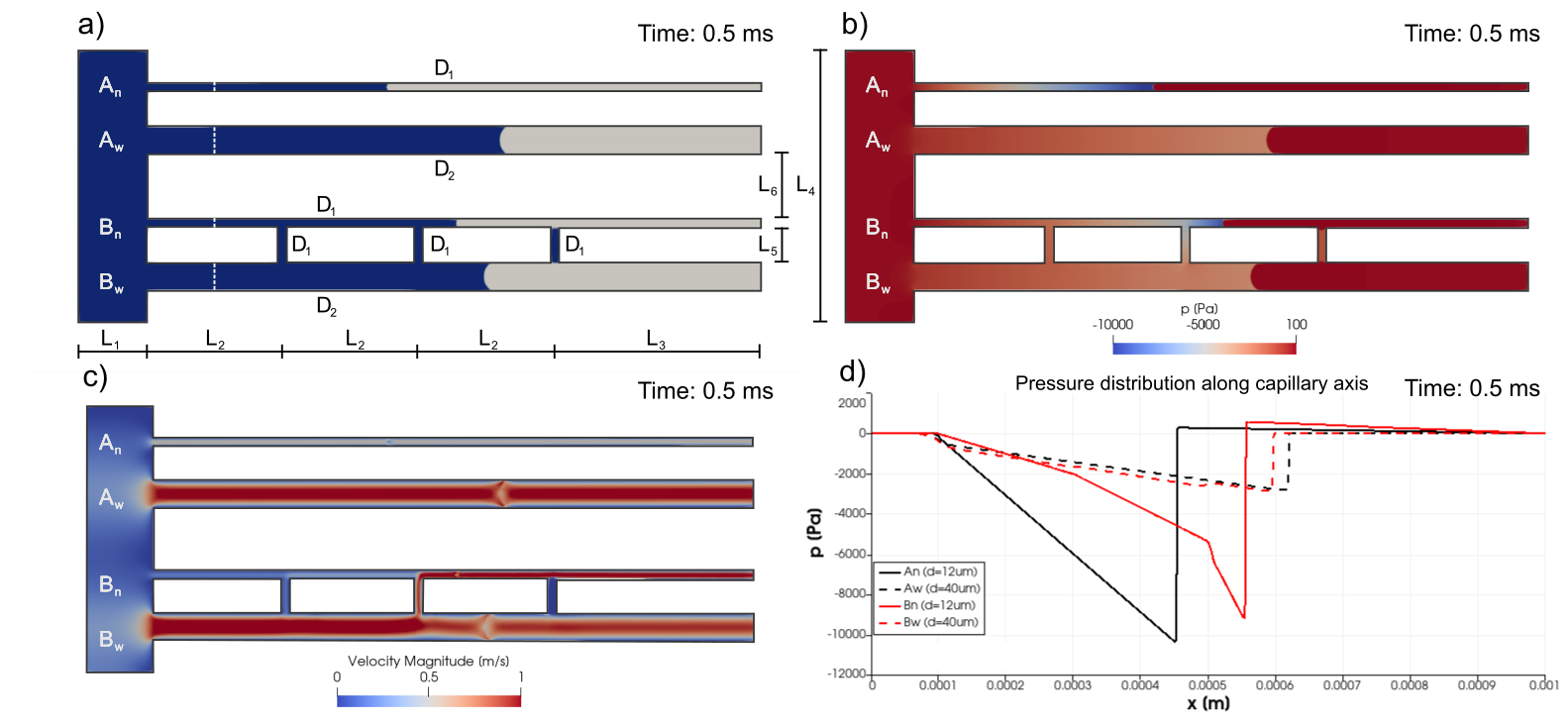}}
\caption{Comparison between independent (set A) and interacting (set B) BCTM demonstrating the cross-flow influence between the capillaries: (a) air-water distribution, (b) pressure distribution, (c) velocity magnitude, and (d) line pressure plots along central lines of
capillaries at t=0.5 ms. Dashed lines represent the air-water interface at t=0 (initial condition), while dimensions from the Figure are specified in Table \ref{tab:sGeom_10_dimensions}. The complete animation showing the full process can be found in the Supplementary Materials.}\label{fig_sGeom_10_0050}
\end{figure*}

Figure \ref{fig_sGeom_10_0050}a shows the geometry and dimensions for numerical simulations. The left part of the domain represents a large pore with four connected capillaries on the right. Those four capillaries are divided into two sets: set A represents independent capillaries, while set B represents interacting capillaries. The subscripts $n$ and $w$ stand for narrow and wide capillaries, respectively. The initial condition for this case is specified as the flat interface on $L_1+L_2/2$ distance from the left edge, and three cross-flow capillaries connecting $B_n$ and $B_w$  are initially filled with water.

\begin{figure*}
\centering{\includegraphics[width=0.95\textwidth]{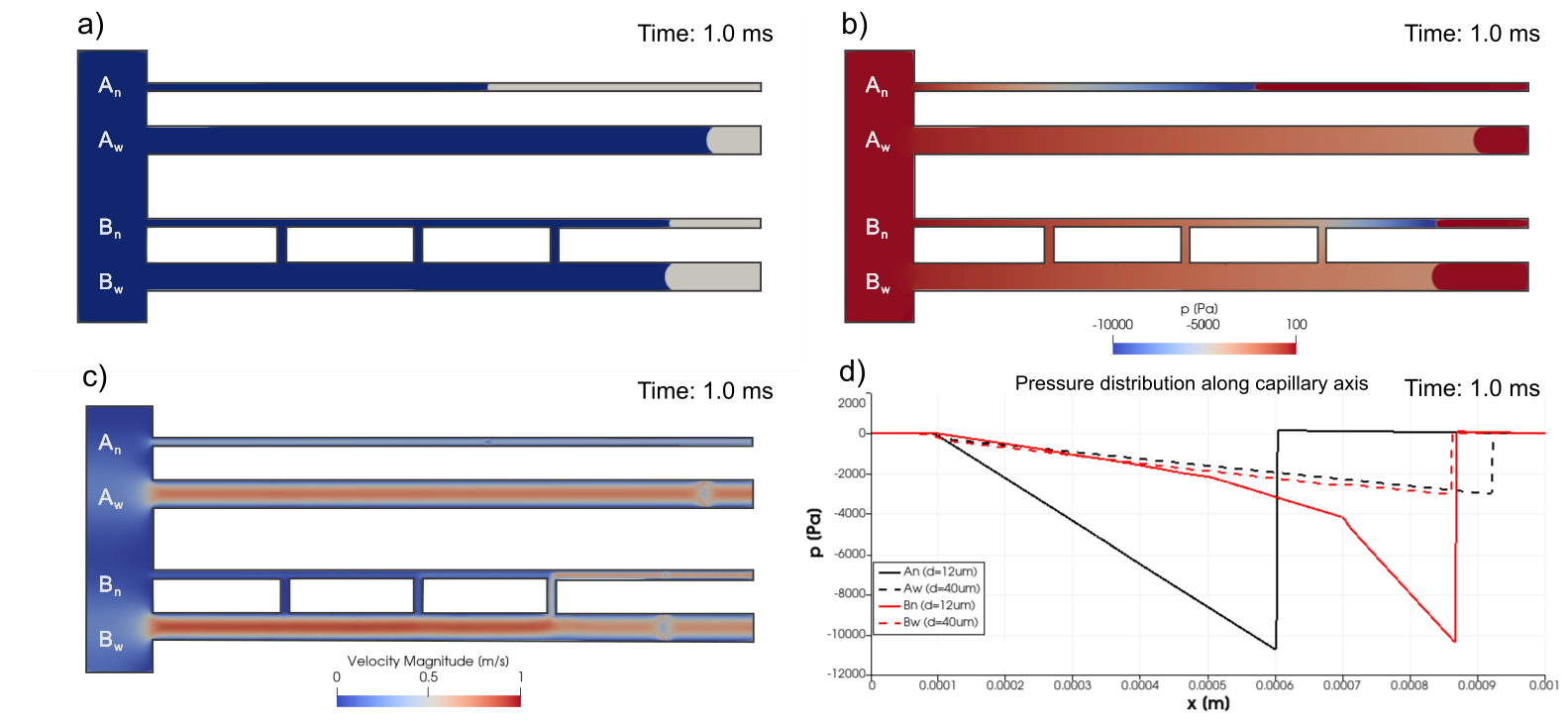}}
\caption{Comparison between independent (set A) and interacting (set B) BCTM demonstrating the cross-flow influence between the capillaries: (a) air-water distribution, (b) pressure distribution, (c) velocity magnitude, and (d) line pressure plots along central lines of
capillaries at t=1.0 ms.}\label{fig_sGeom_10_0100}
\end{figure*}

Figures \ref{fig_sGeom_10_0050} and \ref{fig_sGeom_10_0100} show air-water distribution, pressure distribution, velocity magnitude distribution and line plots of the pressure through the central line passing each capillary at two different times (t=0.5 ms and t=1.0 ms). As expected, the interface position in the non-interacting capillaries (set A) follows the behavior predicted by the Lucas-Washburn equation, i.e. (except in the very beginning due to inertia effects) the wide capillary is filled faster than the narrow one. As previously mentioned, the capillary pressure is larger for narrow capillary (black solid vs black dashed line in Figures \ref{fig_sGeom_10_0050}d and \ref{fig_sGeom_10_0100}d where pressure jumps at the air-water interface represents the value of dynamic capillary pressure $P_c$), however, the friction losses are also larger for narrow capillary which is visible by comparing the slopes of the same pressure lines (black solid vs black dashed). On the other hand, for interacting capillaries (set B) imbibition behavior is different from set A, and interconnections between $B_n$ and $B_w$ affect the flow in such a way that imbibition in the wide capillary $B_w$ is slower (compared to $A_w$), while imbibition in the narrow capillary $B_n$ is faster (compared to $A_n$). The reason for this can be explained by examining pressure and velocity fields and acknowledging that fluid flows from higher toward lower pressure areas.

First, we can notice that larger underpressure is created in the narrow tube ($B_n$) than in the wide tube ($B_w$) due to the smaller radius. Once both meniscus have reached the first cross-flow capillary, pressure differences will lead to cross-flow between two tubes where a part of the flow through $B_w$ will be redirected towards $B_n$. As already mentioned, interacting BCTM assume pressure equilibrium between capillaries containing the same fluid (at the same distance $x$). However, from the Figures \ref{fig_sGeom_10_0050} and \ref{fig_sGeom_10_0100} (obtained by solving full flow equations), it is clear that there is pressure difference between two capillaries. This pressure difference is needed to drive a fluid flow between neighboring tubes and is largest at the latest cross-flow tube where most of the water flow between capillaries occurs (e.g., the difference is visible in 2D pressure plots or by comparing red solid vs red dashed lines in line plots). However, pressure difference and fluid transfer do exist in other areas and not only close to the meniscus (i.e., the latest cross-flow capillary) as generally assumed. The larger the radius of the interconnecting capillary, the smaller the difference in imbibition speed and pressure between $B_w$ and $B_n$ tubes is expected. 

\begin{figure}
\centering{\includegraphics[width=1.0\columnwidth]
{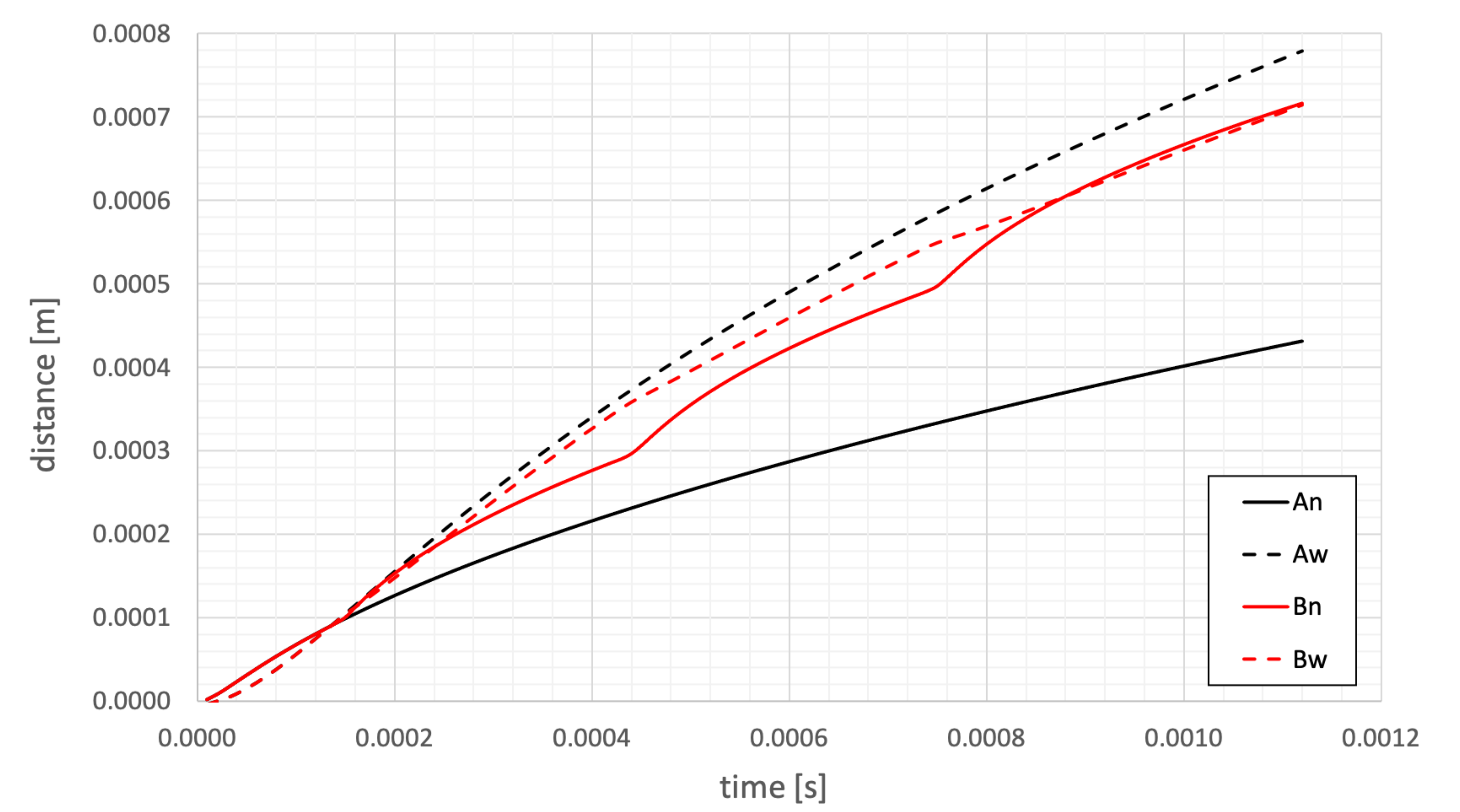}}
\caption{Comparison between independent ($A_n$,$A_w$) and interacting ($B_n$,$B_w$) capillaries: imbibition distance as a function of time.}\label{fig_sGeom_10_x(t)}
\end{figure}

Moreover, Figure \ref{fig_sGeom_10_x(t)} shows the imbibition distance as a function of time for all four capillaries. Here it is interesting to notice that imbibition in narrow tube $B_n$ accelerates whenever it passes a cross-flow capillary, which is again followed by a decrease in imbibition speed as it progresses further away. This speed-up is the outcome of mainly two reasons. First, as soon as the interfaces pass a cross-flow capillary, the large pressure difference between $B_w$ and $B_n$ existing before cross-flow will be reduced by increasing pressure in $B_n$ and reducing in $B_w$ which will initiate temporal acceleration of meniscus in $B_n$. Furthermore, the cross-flow leads to the situation where overall viscous dissipation through the narrow capillary is now significantly lower, especially when the meniscus is close to cross-flow capillary (due to large water-air viscosity ratio) as now significant portion of the flow required for advancing meniscus in $B_n$ is flowing through $B_w$ where friction is significantly lower due to larger radius. This decrease in the pressure gradient in the upstream part of the narrow tube and redirection of fluid from wider to narrower parts of porous media due to the suction effect is one of the reasons why imbibition in narrow capillaries is generally faster, contrary to the prediction of basic Lucas-Washburn equation. 

Additionally, we also notice here that capillary pressure at the air-water interface is different between non-interacting and interacting capillaries (e.g. $A_n$ vs $B_n$ on Figures \ref{fig_sGeom_10_0050}d or \ref{fig_sGeom_10_0100}d). The reasons for this are related to flow dynamics and interface movement and reconfiguration which generally tend to reduce $P_c$. Similar behavior was reported in section \ref{subsec_Detail} during the pore-filling process where differences between dynamic and static (Young-Laplace) capillary pressures were observed.

In this example, three short cross-flow (interconnecting) capillaries that connect $B_n$ and $B_w$ tubes were set as initially filled with water. The main reason for this was to have the direct (transient) comparison with set $A$ of non-interacting capillaries. {The results for initially dry cross-flow capillaries do lead to slightly different imbibition speeds due to cross-flow capillary filling and air flow downstream of the imbibition front; however main conclusions stated here remain the same. Animations for both initially dry and initially filled cross-flow capillaries are included in Supplementary Information}.

Finally, in this example, the interfaces between $B_n$ and $B_w$ were following each other relatively closely (Figure \ref{fig_sGeom_10_x(t)}). As explained, the interface in $B_n$ would accelerate every time it passes a new cross-flow capillary. However, as the front in the narrow capillary would move further away from it, it would slow down due to high viscous dissipation. Thus, for this particular case, the imbibition front of interacting capillaries moves relatively uniformly through the domain. In the following, the same size capillaries will be used; however, the configuration will be changed to demonstrate another reason why imbibition can be significantly faster in narrow capillaries compared to wide ones.

\subsubsection{Influence of narrow contractions on imbibition dynamics and air trapping}
\label{subsec_InfluenceOfThroatsAndAirTrapping}

\begin{figure*}
\centering{\includegraphics[width=0.9\textwidth]{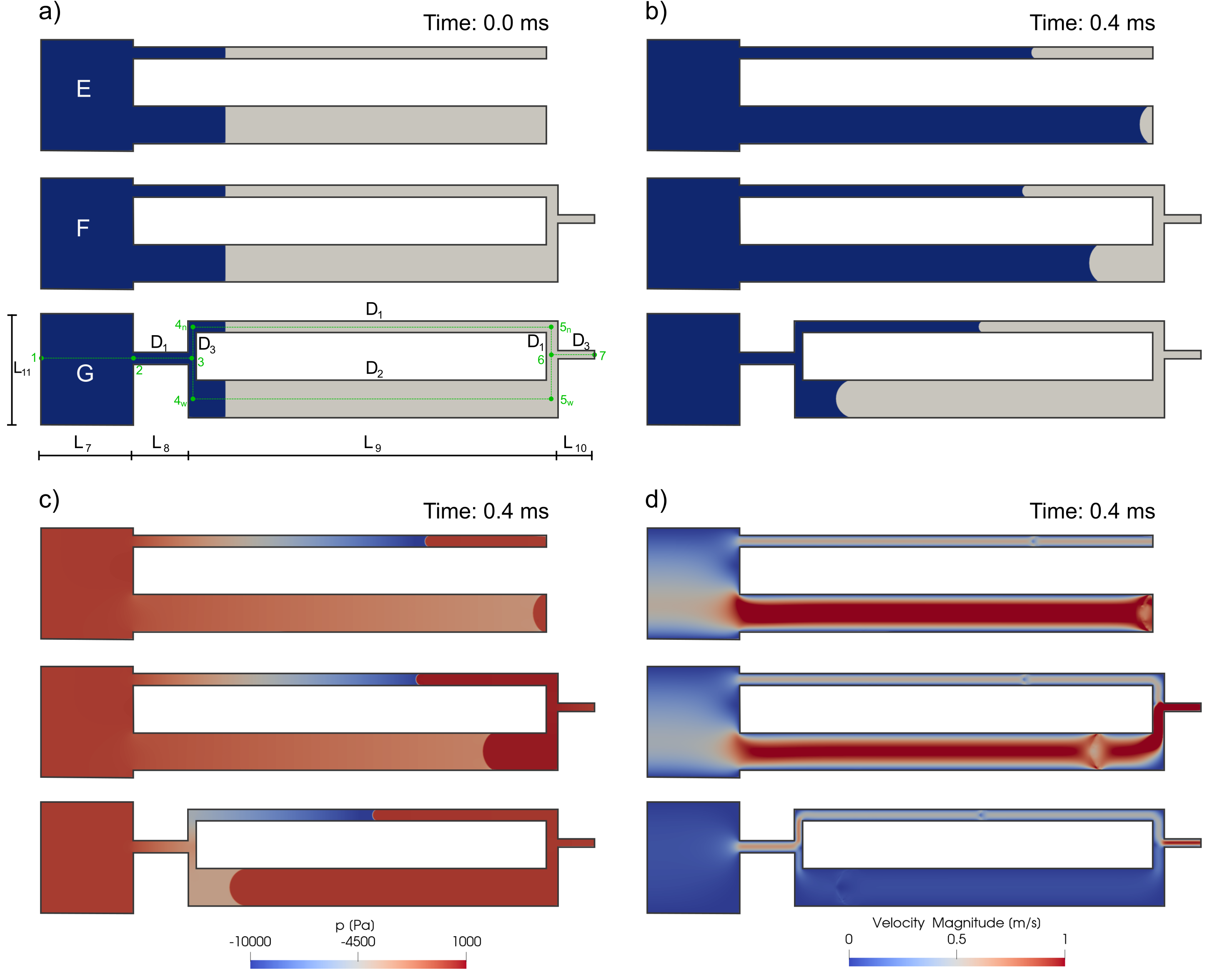}}
\caption{Influence of narrow contractions: (a) air-water initial condition, (b) air-water distribution, (c) pressure distribution and (d) velocity magnitude at t=0.4ms. Set E follows basic Lucas-Washburn behavior, while set F is slower due to airflow through the downstream contraction, demonstrating the influence of air viscosity. Upstream contraction in set G leads to totally different behavior and shows that narrow contractions could have a larger influence on imbibition than cross-flow between parallel capillaries, as shown in the previous example. Dimensions denoted in Figure \ref{fig_sGeom_10_throats_01}a are specified in Table \ref{tab:sGeom_10_dimensions}. The complete animation showing the full process can be found in the Supplementary Materials.}\label{fig_sGeom_10_throats_01}
\end{figure*}

In the following section, we look into the influence of contractions (i.e. narrow throats) on imbibition dynamics and the trapping of the non-wetting phase. We compare the set of independent tubes (set $E$, the same as set $A$ from the previous example) to the set $F$ containing narrow throat at outlet boundary and the set $G$ containing narrow throats both at inlet and outlet as shown in Figure \ref{fig_sGeom_10_throats_01}a. 

First, we compare differences between case $E$ and case $F$, i.e. we look into the influence of the narrow throat at the outlet. Figure \ref{fig_sGeom_10_throats_01}b shows the interface position at t=0.4 ms, just before the interface of the fastest capillary ($E_w$) has reached the end of the domain. Like in the previous example, we use subscripts $w$ and $n$ to distinguish between wide and narrow capillaries, respectively. We can see that set $F$ has the same behavior as a set $E$ (following Lucas-Washburn prediction) where larger capillary is filled first; however, we can also see that filling is slower than in the case of independent capillaries (set $E$). The only difference here is a narrow throat at the end of the domain, where neither of the air-water interfaces in $F_w$ or $F_n$ didn't reach the throat yet. Thus, this behavior has to be related to the flow of the air through the narrow throat at the end of the domain, which can be visible by looking at the pressure and velocity distribution in Figures \ref{fig_sGeom_10_throats_01}c and \ref{fig_sGeom_10_throats_01}d. We can see that, compared to set $E$, the pressure in the air phase (downstream/right from the interface) is higher (darker red color) which slows down the imbibition of water in set $F$. The reason for this can be understood by looking at a velocity magnitude field where the local velocity through the outlet throat is very high (because the sum of discharges from both $F_w$ and $F_n$ needs to flow through this narrow contraction) which needs a non-negligible pressure gradient in the gas phase even considering the small viscosity of the air. As the outlet pressure is set the same as at the inlet pressure (i.e. relative atmospheric pressure p=0), this leads to increased air pressure downstream of the interface acting as an opposing force and slows down water imbibition. 
The air viscosity is often neglected since water is significantly more viscous fluid (approximately 55 times more viscous than air). However, here we see that air viscosity can play a role and slow down overall imbibition. This is expected to have the most significant effect in the early stages of the imbibition (where local velocities are largest) and for the case where porosity is reducing downstream compared to porosity at the imbibition front.

\begin{figure}
\centering{\includegraphics[width=1.0\columnwidth]
{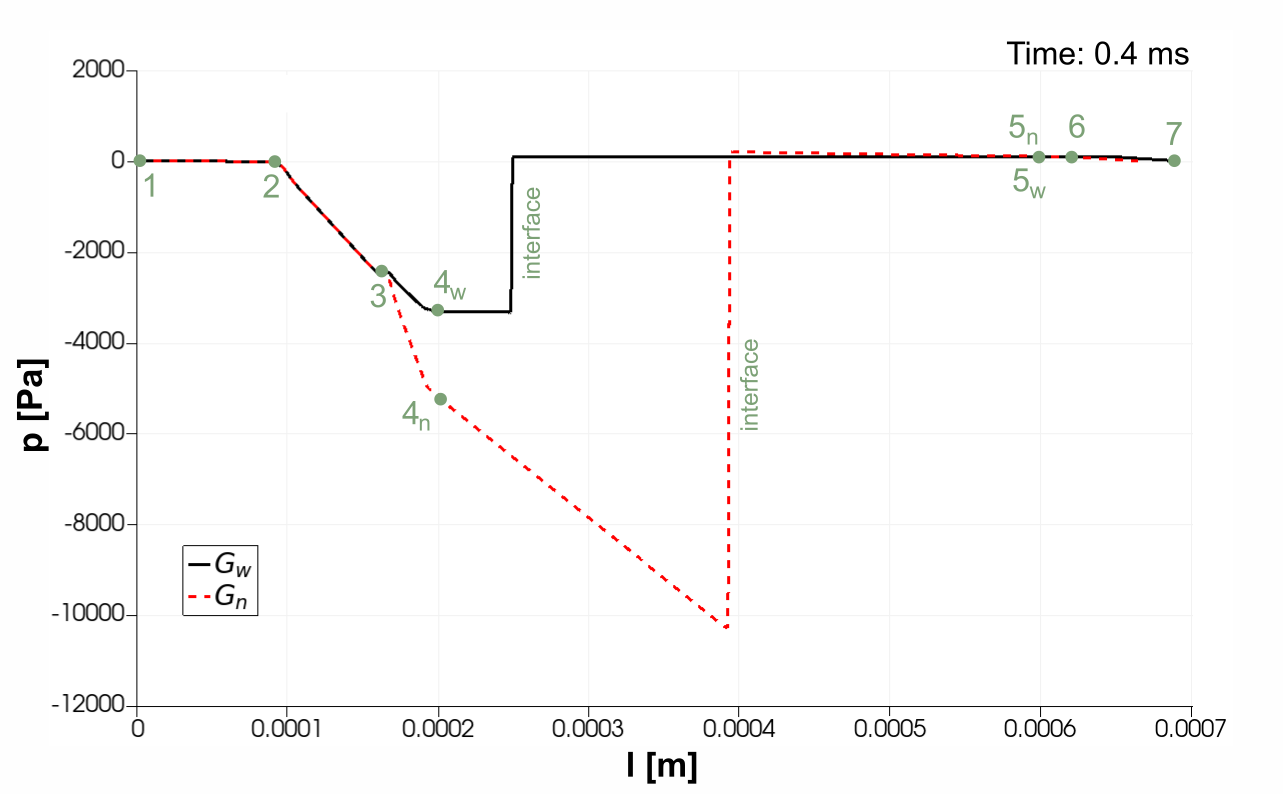}}
\caption{Pressure
distribution for G set at 0.4ms along poly-lines denoted at Figure \ref{fig_sGeom_10_throats_01}\label{fig_sGeom_10_throats_polyLine}a}
\end{figure}

However, the effect of the narrow contraction will become more significant and have a larger effect on overall imbibition dynamics once it is filled by water, as seen in the results of the set $G$, which has additional contraction between the inlet pore and two capillaries. When compared to two previous sets, imbibition dynamics in set $G$ is not just significantly slower but now has an opposite behavior as a smaller tube is filled faster than a larger one (Figures \ref{fig_sGeom_10_throats_01}b). This behavior is often seen in real porous media and can be explained by the pressure and velocity distributions in Figures \ref{fig_sGeom_10_throats_01}c and \ref{fig_sGeom_10_throats_01}d. A particular feature in this example is the T-junction where the flow is divided from the inlet throat (connected to a large pore) to wide and narrow ($G_w$ and $G_n$) capillaries. We can see that most of the flow is redirected toward narrow capillary $G_n$ and the reason for this is the local pressure distribution. To get better insight, Figure \ref{fig_sGeom_10_throats_polyLine} shows pressure distribution along two polylines passing centers of $G_n$ and $G_w$ capillaries at time $t=0.4$ ms (polylines are denoted with numbers and green dashed lines in Figure \ref{fig_sGeom_10_throats_01}a). Point $3$ is the location where the flow starts branching into two directions and most of the flow is redirected toward $G_n$ because of lower pressure at point $4_n$ than at $4_w$. This larger under-pressure at $4_n$ results from higher capillary pressure $P_c$ at the air-water interface because of the smaller radius of the narrower tube. As previously discussed and demonstrated by the results of non-interacting capillaries, imbibition in narrow tube will create larger under-pressure but, due to larger viscous dissipation, overall imbibition will be slower than in the wider capillary as predicted by the Lucas-Washburn equation. In the case of set $G$, viscous losses are higher in the narrow tube (slope of the red dashed line between point $4_n$ and interface position) than in the wider tube (slope of the solid black line between point $4_w$ and interface). However, in this case, the difference is that capillaries are not connected to large pore as before, but rather source of inflow water is limited by narrow contraction (throat) and the local pressure distribution in the T-junction redirects most of the flow toward the narrow capillary.

\begin{figure*}
\centering{\includegraphics[width=0.9\textwidth]{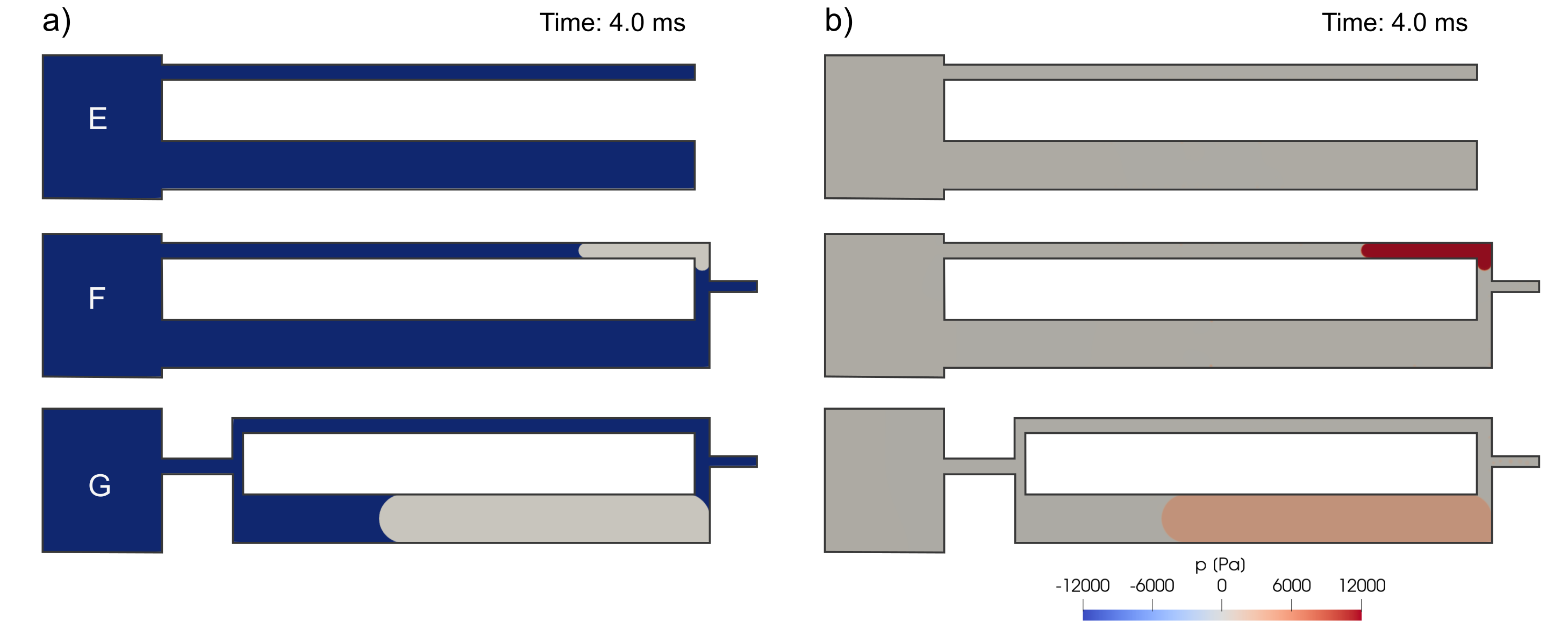}}
\caption{Influence of contractions on entrapment of the air (non-wetting) phase: (a) air-water distribution and (b) pressure distribution after steady-state conditions have been reached.}\label{fig_sGeom_10_throats_03}
\end{figure*}

Moreover, because two capillaries with different imbibition speeds merge into the narrow throat at the outlet of the domain (for cases $F$ and $G$), this will result in the entrapment of the non-wetting phase in the system. Figure \ref{fig_sGeom_10_throats_03} shows air-water distribution and pressure fields for all three cases at the time t=4 ms where steady-state configurations have been reached. For case $E$, there is no air entrapment as there is no merging of capillaries. 
For the middle case $F$, imbibition in the wide capillary will reach the outlet throat sooner than narrow capillary which will lead to trapping of air in the narrow capillary $F_n$ while for the case $G$ situation will be the opposite and air will be entrapped in wide capillary $G_w$ as shown in the Figure \ref{fig_sGeom_10_throats_03}a. Figure \ref{fig_sGeom_10_throats_03}b shows corresponding pressure fields where it is clear that the trapped air phase has a higher pressure than water as is expected for the non-wetting phase. Since there is no flow after the steady state is reached, the value of pressure in the water phase is constant and defined by boundary conditions (i.e. $p=0$). Thus, pressure in the air phase is the same (to the accuracy of a numerical algorithm) as the static $P_c$ calculated using the Young-Laplace equation. 

Finally, we see that the final positions and the volumes of the entrapped air between different cases are significantly different, which can have an important effect on many physical and chemical processes for transport in porous media. Moreover, note that the downstream air-water interface is not located directly at the outlet throat position. For the case $F$, this is mostly due to inertia effects as the interface entering a narrow tube from a wide capillary will have significant momentum and push trapped air back downstream until viscous forces dissipate this momentum. For the case $G$, the reason for this is not inertia but rather unbalanced capillary pressures between downstream and upstream air-water interfaces, pushing trapped air volume upstream until curvature on both ends becomes the same and pressures in the system equilibrate. We refer to the animations provided in the Supplementary Information for additional insight into the described processes.

%% file: 04_Implications.tex
\section{Implications}\label{sec_implications}

Understanding spontaneous imbibition phenomena from the pore-scale perspective is of fundamental importance and has large potential to improve the predictive capabilities of many simplified models. The DNS approach enables detailed microscopic insight into multiphase flow physics by solving full flow variables without simplifications. This work combined DNS with simplified 2D geometries to simulate different pore-level processes relevant during spontaneous imbibition.

First, dynamic effects in the transition zone from throat-to-pore and pore-to-throat were investigated. Such events are parts of every natural porous media and are generally implemented in PNM as simple capillary pressure rules. This study confirms that throat-filling is a fast process, and thus, dynamic effects in the transition zone are not expected to influence the imbibition prediction by PNM. On the other hand, interfaces passing the transition zone during a pore-filling event (a situation before and after positive capillary pressure is established across the whole meniscus) can significantly slow down imbibition. In the case of spontaneous imbibition with sharp changes in geometry, the inertial forces are crucial and determine the time needed for the interface to pass the transition zone. The larger water momentum at the transition zone's entrance means a faster filling process. When momentum is insufficient to overcome abrupt changes in the shape of pore space, the capillary barrier will occur, and imbibition into the pore-body will be prevented. This is particularly significant on smaller scales (due to smaller momentum), which generally contradicts the general hypothesis used in conventional PNM that smaller pores are filled first due to larger capillary pressure. The inertial effects are often considered irrelevant in porous media due to low (averaged) velocities. However, from the pore-scale perspective, it is clear that inertia significantly influences the local interface reconfigurations, which eventually can lead to significantly different imbibition speeds and invasion patterns in complex porous media. 

Moreover, the apparent contact angle is generally lower than its static value during the imbibition process. 
This is true even for a single capillary with a constant cross-section, and dynamic contact angle approaches have been developed to describe such behavior \citep{bianchi_janetti_effect_2022,zhangMathematicalModelTwoPhase2023}. However, this phenomenon is naturally included in the VOF framework, even when a static (constant) contact angle is specified as a boundary condition. This is because the interface is advected based on the multidimensional pressure and velocity fields obtained by solving the full mass and momentum equations. This further leads to dynamic capillary pressure not being constant across the meniscus. VOF simulations show that capillary pressure has the lowest value in the middle of the meniscus, which can be significantly lower than predicted by the Young-Laplace (Y-L) equation, and its value increases by approaching wall boundaries, where it can significantly exceed the Y-L value. Such insight not only enhances our fundamental understanding but could also contribute to the development of improved dynamic contact angle models.

Once the interface reaches the entrance of the pore body, a complex meniscus reconfiguration, with a whole range of capillary pressure values, will occur. This will be followed by multiple dynamical transitions between concave and convex curvatures of different interfacial zones. Even after interfaces pass the transition zone, oscillations in the pore-body can take some time until the meniscus reaches steady-state (dynamic) curvature again. As suggested by \citet{ferrari_inertial_2014}, these irreversible reconfigurations characterize the amount of surface energy that is spontaneously released and transformed into kinetic energy and can be related to the pressure drop in Darcy’s law. Thus, more work is needed to better understand the relevance and upscaling of such pore-scale reconfigurations.

Furthermore, we presented a DNS comparison between non-interacting and interacting bundles of capillary tube setup where cross-flow exchange occurs perpendicularly to the flow direction. This was followed by investigating the influence of narrow contractions, where the flow from a single capillary divides into multiple capillaries. Both cross-flow and narrow contraction can explain why imbibition is generally faster into narrow capillaries (opposite to predictions of the basic Lucas-Washburn equation). However, our results suggest that narrow contractions could have a far more significant effect than cross-flow between capillaries. Moreover, narrow contractions, which are generally not included in BCTM, can explain why the non-wetting phase is generally trapped in large pores, which has a significant influence on many different applications. An example is the trapping of air in the capillary movement of water through concrete, where the air plays an important role in the context of durability (e.g., freeze-thaw damage or reinforcing steel corrosion).

In this work, incompressible flow assumption was used. Such an assumption is appropriate in most porous media applications since the flow of a compressible fluid (such as air) at a low Mach number ($\sim$ Ma < 0.3) is essentially incompressible \citep{ferzigerComputationalMethodsFluid2020} (note that incompressible \textit{flow} and incompressible \textit{fluid} are different terms). However, once the air phase gets trapped, compressibility effects could become important since over-pressure in the air phase (imposed by capillary pressure) could potentially lead to air compression and dissolution \citep{smithDissolutionKineticsTrapped2020}. While this topic extends beyond the scope of the present study, it is worth mentioning that such effects are anticipated to be particularly pronounced on smaller spatial scales (due to the significant capillary pressure), as well as on longer temporal scales. Considering that the provided example had a maximum capillary pressure of around 0.12 bar and a full simulation time of 4 ms, these effects are not expected to influence conclusions from this study. However, the long-term dissolution of trapped air has important consequences for many applications and could be one of the explanations for the later stage of (anomalous) water absorption into cementitious materials \citet{zhangDifferentAnomaliesTwostage2024}.

We mostly discussed how investigated processes are implemented or how gained insight could be used for simplified pore-scale models (PNM and BCTM). However, all the mentioned processes are relevant to imbibition physics; thus, macroscale models based on a homogenization approach (i.e., continuum-based models using Darcy law) could also benefit from such findings. Finding a way to upscale described pore-level physics, either by modifying governing equations or improving the description of material properties (e.g., relative permeability curves), has the potential to improve their predictive capabilities.

%% file: 05_Conclusions.tex
\section{Conclusions}\label{sec_conclusions}

Understanding how water is absorbed in dry porous materials is crucial due to its wide range of applications. It's important to consistently enhance our knowledge and ability to predict these processes. In this work, accurate physical simulations of air-water multiphase flow have been performed using direct numerical simulations (DNS) to investigate different pore-scale effects and their impacts on imbibition dynamics during capillary-driven water imbibition. The main findings can be summarised as follows:

\begin{itemize} 
    \item Throat-filling process is a more energetically favorable process than pore-filling. This can lead to significantly increased time for the meniscus to pass the transition zone (from throat to pore body) during a pore-filling event. Simple capillary pressure rules implemented in PNM do not account for such complex dynamic effects.
    \item Dynamic capillary pressure $P_c$ is generally not constant across the interface and can differ significantly from its equilibrium value predicted by the Young-Laplace equation. This leads to uneven meniscus curvature with a different (apparent) contact angle than its intrinsic (local) value, even for the flat and homogeneous surface. These effects are inherently included in the VOF method, even when static contact angle boundary condition is used.  
    \item Inertia plays an important role during pore-filling events and can be a deciding factor if pore-filling will occur or not. Larger water momentum at the entrance of the pore body leads to faster interface reconfigurations and, eventually, faster pore filling. If the momentum is insufficient, the capillary barrier can occur, and imbibition will be prevented. This is particularly significant on the smaller scales (e.g. submicron porosity), which generally contradicts the general hypothesis used in conventional PNM that smaller pores are filled first due to larger capillary pressure.
    \item The pressure equilibrium assumption used in interacting BCTM, which assumes that pressure drop and fluid transfer occur only at the meniscus locations, is generally not valid; rather, a more complex hydrodynamic situation exists in reality, and the potential significance of these effects should be considered and further investigated when applying BCTM.
    \item Narrow contractions where flow divides to multiple capillaries could be a more important reason why imbibition is generally faster in narrow capillaries than the cross-flow between capillaries. 
    \item Narrow contractions are important components for non-wetting phase entrapment and can explain why the non-wetting is often trapped in the large pores. Including such effects has the potential to improve the predictive capabilities of existing BCTM.  
    \item The viscosity of the air, in some cases, can have a non-negligible effect and slow down overall imbibition.
    
\end{itemize}

The findings from this study have the potential to offer valuable insights to researchers seeking to advance their understanding of imbibition processes. Furthermore, this study could be particularly beneficial to those working with simplified pore-scale models like PNM and BCTM.  
Additionally, it illustrates the utility of DNS in answering critical questions, examining different hypotheses, and validating different components of existing predictive models.